\documentclass[twocolumn]{aastex701}

\usepackage{graphicx}	
\usepackage{amsmath}	
\usepackage{amssymb}	
\usepackage{threeparttable}
\usepackage{comment}
\usepackage{caption}  
\usepackage{subcaption}

\def\chandra    {{\em Chandra}\/}
\def\cha    {{\em Chandra}\/}

\def\XMM        {{\em XMM}\/}
\def\XMM        {{\em XMM-Newton}\/}

\begin{document}

\title{The shocking features in the closest rich galaxy cluster Norma}

\author[orcid=0000-0003-0628-5118]{Chong Ge}
\affiliation{Department of Astronomy, Xiamen University, Xiamen, Fujian 361005, China}
\email[show]{chongge@xmu.edu.cn}  

\author[orcid=0000-0001-5880-0703]{Ming Sun} 
\affiliation{Department of Physics \& Astronomy, University of Alabama in Huntsville, 301 Sparkman Dr NW, Huntsville, AL 35899, USA}
\email{fakeemail1@google.com}

\author[orcid=0000-0003-0231-3249]{Mpati Ramatsoku}
\affiliation{Centre for Radio Astronomy Techniques and Technologies (RATT), Department of Physics and Electronics, Rhodes University, Makhanda, 6140, South Africa}
\affiliation{INAF-Osservatorio Astronomico di Cagliari, Via della Scienza 5, I-09047 Selargius (CA), Italy}
\email{fakeemail2@google.com}

\author[orcid=0000-0002-0775-6017]{Chris Nolting}
\affiliation{Gustavus Adolphus College, Saint Peter, Minnesota, USA}
\email{fakeemail4@google.com}

\author[orcid=0000-0003-4351-993X]{B\"{a}rbel S. Koribalski}
\affiliation{Australia Telescope National Facility, CSIRO, Space and Astronomy, P.O. Box 76, Epping, NSW 1710, Australia}
\affiliation{School of Science, Western Sydney University, Locked Bag 1797, Penrith, NSW 2751, Australia}
\email{fakeemail5@google.com}

\begin{abstract}

The merger shocks generated by the collision of galaxy clusters elevate the pressure within the intracluster medium, significantly influencing the evolution of embedded cluster galaxies.
We detect a merger shock (Mach number $\sim 1.3$) on the northwest side of the closest rich galaxy cluster Norma (A3627), using \XMM\ and \cha\ data.
The textbook ram pressure stripping (RPS) galaxy ESO~137-001 appears to be located in the post-shock region. The shock boosts RPS and may induce the formation of the brightest known X-ray tail behind a cluster late-type galaxy.
Another prominent head-tail radio galaxy ESO~137-007, with one of the longest radio continuum tails ($> 500$ kpc), is also likely in the post-shock region. The shock may have reversed the upstream jet to a one-sided radio head-tail morphology. Moreover, the shock can strip and roll the jet cocoon into a vortex ring structure like a `smoke ring' behind the end of the jet as observed by the ASKAP data. Therefore, the cluster merger shock can remarkably change cluster galaxies.
Furthermore, Norma is the second brightest non-cool-core cluster following the Coma cluster, with a cool core remnant on its southeast side. Its original cool core may be disrupted by cluster mergers and/or active galactic nuclei.

\end{abstract}


\section{Introduction}

Galaxy clusters form hierarchically through the merger and accretion of smaller substructures at the intersection of the cosmic web. Merging clusters with induced shocks can affect the evolution of cluster member galaxies. 
The merger shocks increase the pressure of the intracluster medium (ICM), in which member galaxies are embedded. As a result, both star formation and activity in active galactic nuclei (AGNs) can be triggered in gas-rich members as a result of the strong compression of their cold gas by the increased pressure (e.g. \citealt{Bekki2010, Roediger2014}).
Boosted AGN and star formation activity have been observed in merging clusters hosting shocks (e.g. \citealt{Umeda2004,Sobral2015,Stroe2021}).
Moreover, member galaxies moving through the ICM can lose gas via ram pressure stripping (RPS; \citealt{Gunn1972,Boselli2022}).
This stripped gas forms a tail behind the galaxy, and some stripped gas can even turn into stars (e.g. \citealt{cramer2019,Ramatsoku2019}). We call such kinds of galaxies as ``RPS galaxies".
The merger shocks can enhance the RPS and star formation in the tails of RPS galaxies, because of the high-pressure shocked ICM (e.g. \citealt{Owers2012,Li2023}).

\begin{figure*}
    \centering  
    \begin{minipage}{0.66\textwidth}  
        \centering  
        \includegraphics[width=\linewidth]{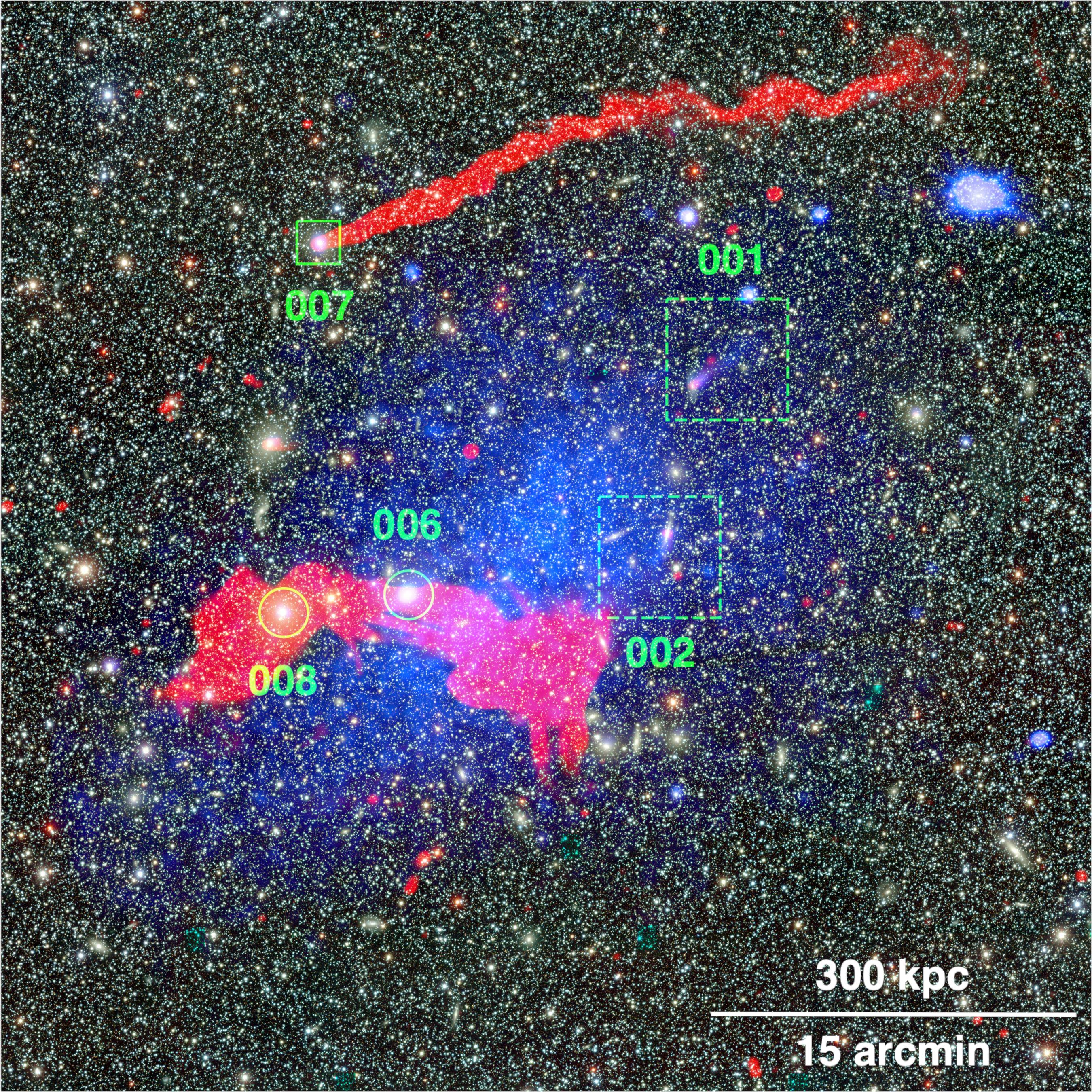}  
    \end{minipage} 
    \begin{minipage}{0.3285\textwidth}  
        \centering  
        \begin{subfigure}{\linewidth}  
            \centering  
            \includegraphics[width=\linewidth]{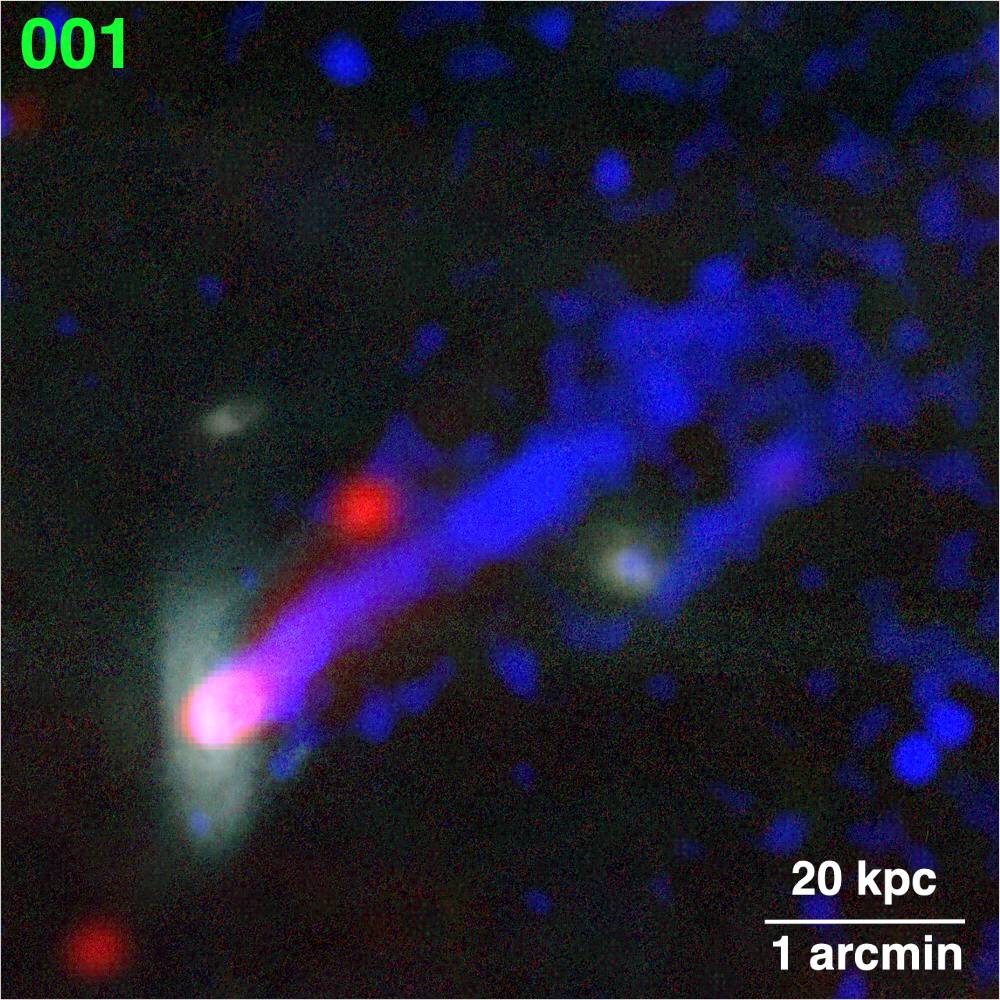} 
        \end{subfigure}\\
        \begin{subfigure}{\linewidth}  
            \centering  
            \includegraphics[width=\linewidth]{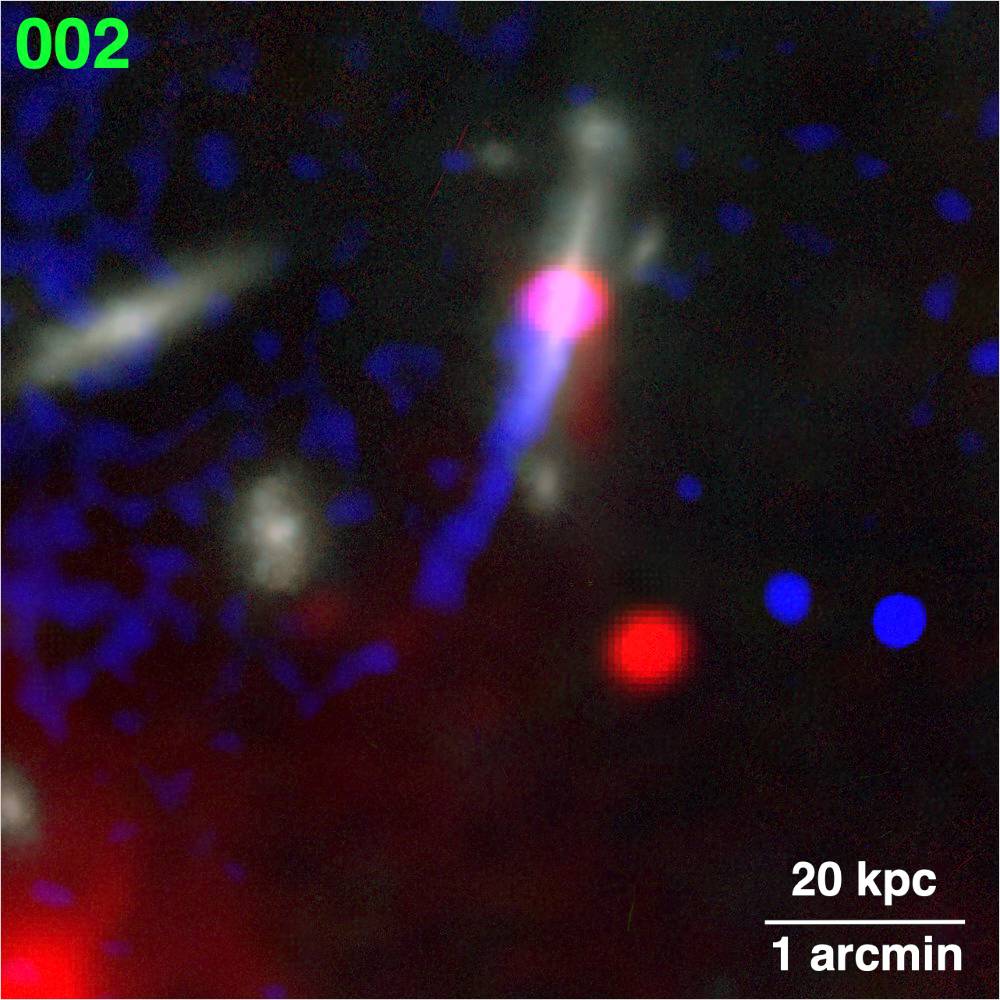}
        \end{subfigure}  
    \end{minipage}  
    
\caption{{\it Left}: optical background image from the Dark Energy Camera Legacy Survey (DECaLS; \citealt{Dey2019}) in the $g, r, z$ bands, overlaid with a red radio image from ASKAP 944 MHz \citep{Koribalski2024}, and a blue X-ray image from the \XMM\ 0.5-2 keV mosaic. The green circles mark two BCGs, ESO~137-006 and ESO~137-008, the green box marks the head-tail radio galaxy ESO~137-007, and the two green dashed boxes mark two RPS galaxies, ESO~137-001 and ESO~137-002, which are enlarged in the right panels.
{\it Right}: zoom-in images of ESO~137-001 and ESO~137-002, with stars removed from the background DECaLS images using {\sc StarNet}\footnote{\url{https://www.starnetastro.com/}}
, overlaid with the red radio image from ASKAP 944 MHz, and the blue X-ray image from the \cha\ 0.5-2 keV residual image.}
\label{fig:rgb}  
\end{figure*} 

The Norma cluster (A3627; Fig.~\ref{fig:rgb}) is the closest massive cluster ($z = 0.0163$, $M\sim10^{15}M_\odot$), and rivals Coma and Perseus in mass and galaxy content (\citealt{Kraan-Korteweg1996}).
However, it attracts less attention than other nearby luminous clusters due to its proximity to the Galactic plane ($b\sim -7^{\circ}$).
A3627 is located at the gravitational center of the Great Attractor and the local Laniakea supercluster, which strongly dominates the growth of the local Universe (e.g. \citealt{Woudt2008,Tully2014}).
A major cluster merger is implied by earlier X-ray observations from the {\em ROSAT}, {\em ASCA}, and {\em Suzaku} that reveal a southeast-northwest (SE-NW) cluster elongation (Fig.~\ref{fig:img}) and a temperature gradient ($5-7$ keV) in the same direction (\citealt{Boehringer1996,Tamura1998,Nishino2012}). 
The merger activity is further supported by the large difference in velocity of two brightest cluster galaxies (BCGs; ESO~137-006 at $\sim$ 5441 km s$^{-1}$ and ESO~137-008 at $\sim$ 3839 km s$^{-1}$) relative to the mean velocity of the cluster members ($\sim$ 4871 km s$^{-1}$; \citealt{Woudt2008}).
One of the BCGs, ESO~137-006, is a textbook example of a wide-angle-tail (WAT) radio galaxy (Fig.~\ref{fig:rgb}), and is one of the brightest radio galaxies in the southern sky (e.g. \citealt{Jones1996,Ramatsoku2020}).
Another prominent radio galaxy is ESO~137-007, which is a head-tail radio galaxy with a tail length of $> 500$ kpc (Fig.~\ref{fig:rgb}), and this is one of the longest radio continuum tails (e.g. \citealt{Jones1996,Koribalski2024}).  
A3627 also hosts two spectacular X-ray/H$\alpha$ tails behind RPS galaxies: ESO~137-001 ($>$ 80 kpc; \citealt{Sun2010}) and ESO~137-002 ($\sim$ 40 kpc; \citealt{Sun2010}; \citealt{Zhang2013}). 
Therefore, A3627 is in the process of an aggressive cluster merger, associated with the formation of a large-scale structure and the violent activity of the member galaxies.

Here we present the discovery of a merger shock on the NW side of A3627. We also report evidence of shock interactions with member galaxies in A3627.
Section~\ref{sec:data} presents the \cha\ and \XMM\ data reduction. Section~\ref{sec:result} reports the X-ray properties of the cluster and merger shock. Section~\ref{sec:discussion} is the discussion. We present our conclusions in Section~\ref{sec:conclusion}.
We assume a cosmology with $H_0$ = 70 km s$^{-1}$ Mpc$^{-1}$, $\Omega_m=0.3$, and $\Omega_{\Lambda}= 0.7$. At A3627's redshift of $z=0.0163$, $1^{\prime\prime}=0.332$ kpc.

\section{Data analysis}
\label{sec:data}
\begin{figure*}
\begin{center}
\centering
\includegraphics[angle=0,width=0.49\textwidth]{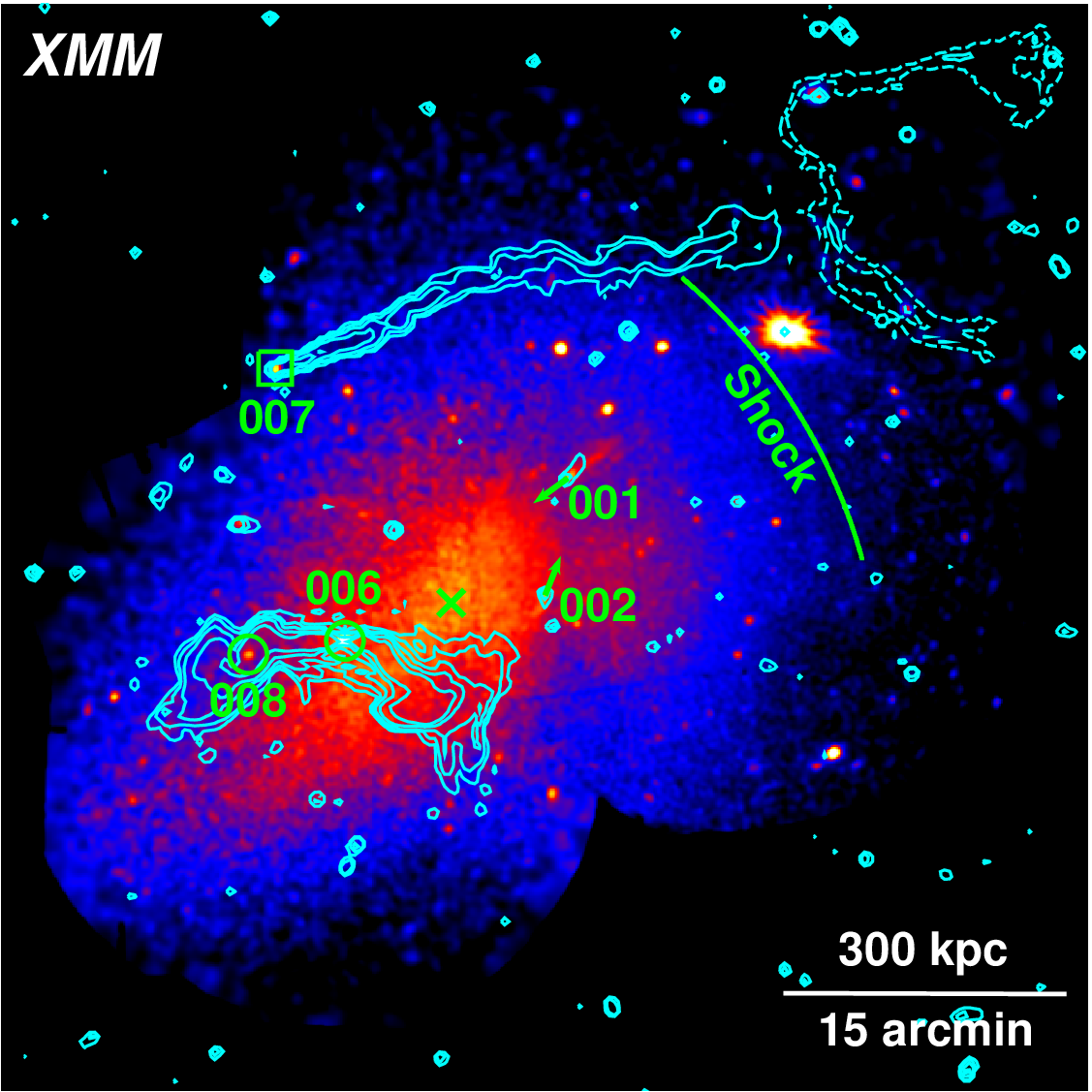}
\includegraphics[angle=0,width=0.49\textwidth]{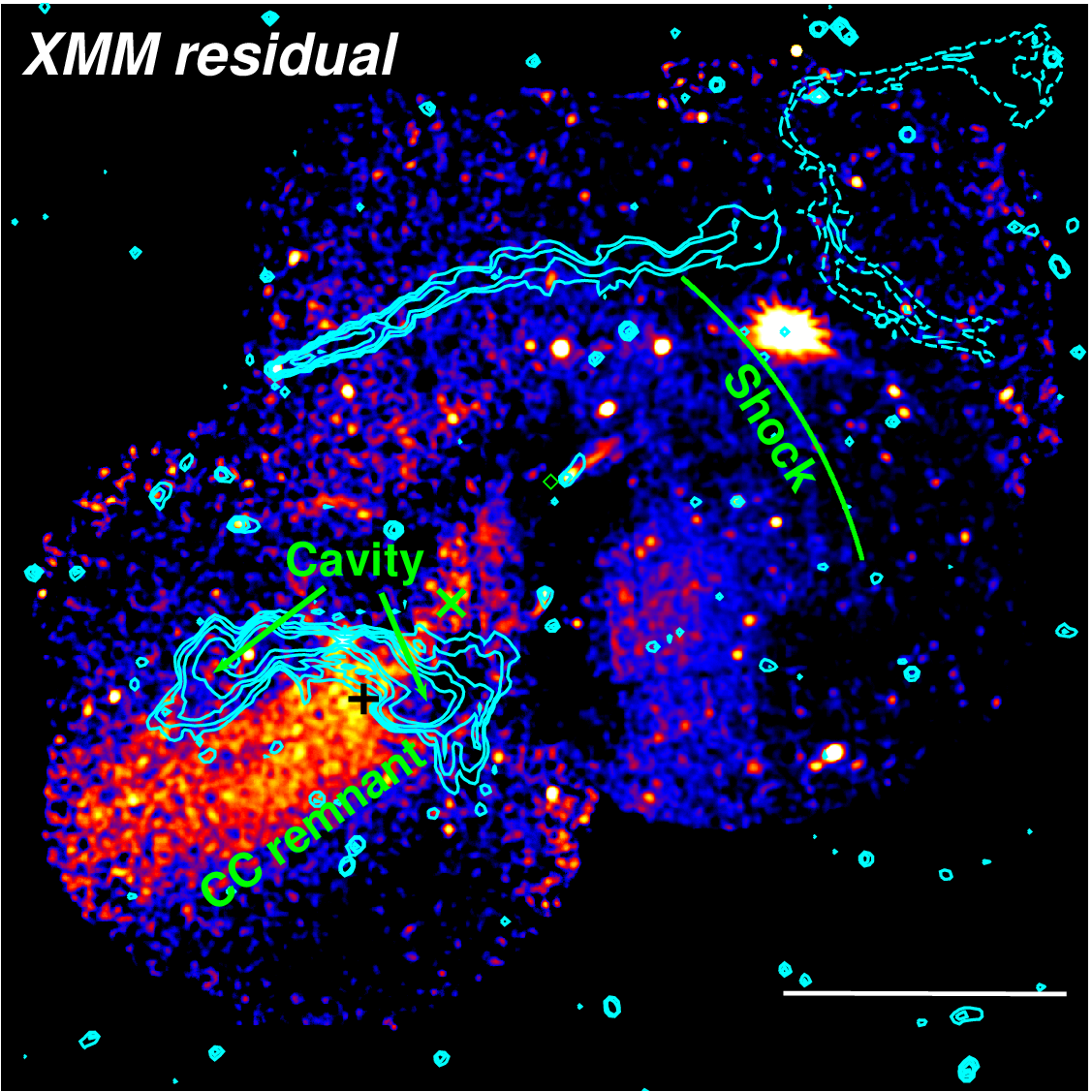}
\includegraphics[angle=0,width=0.49\textwidth]{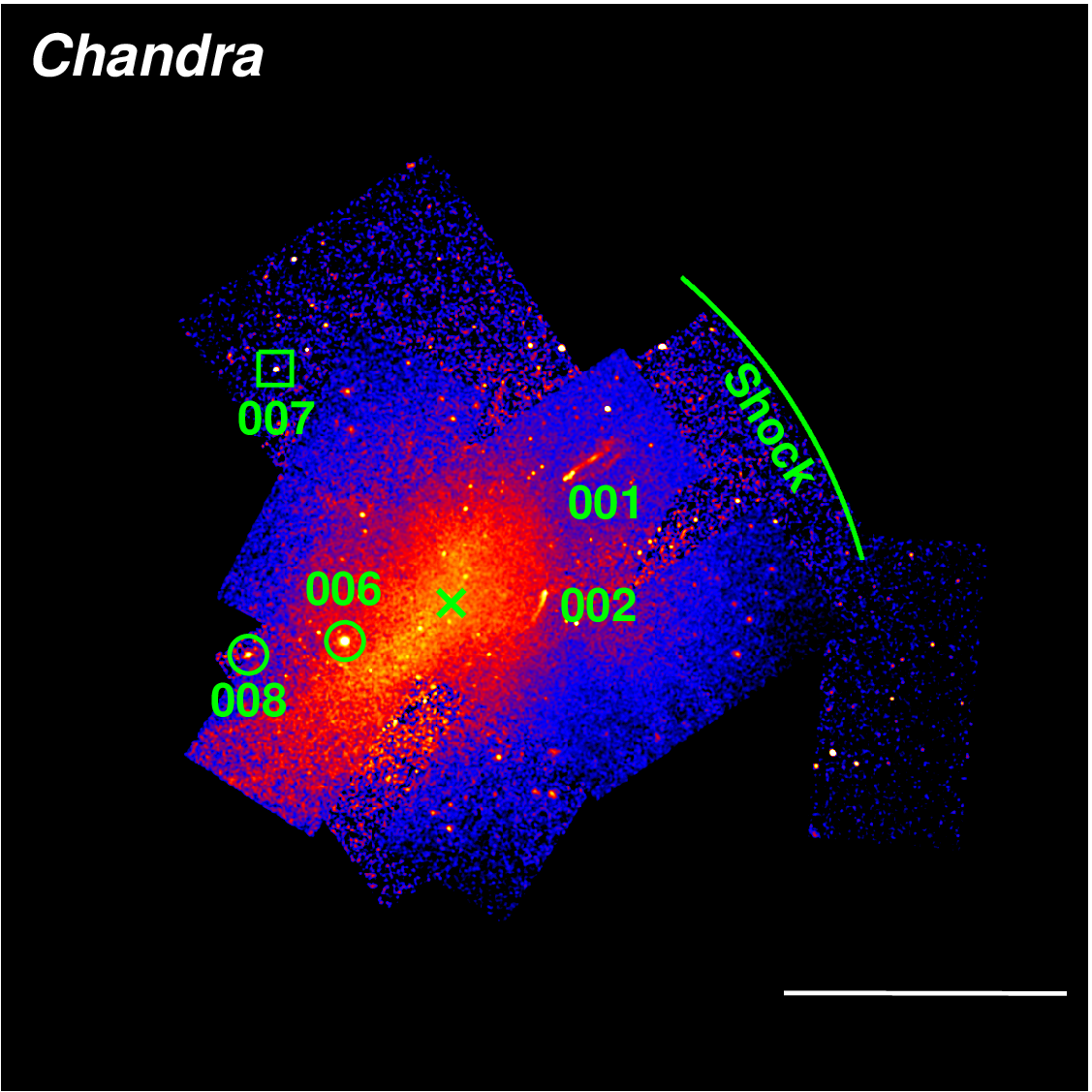}
\includegraphics[angle=0,width=0.49\textwidth]{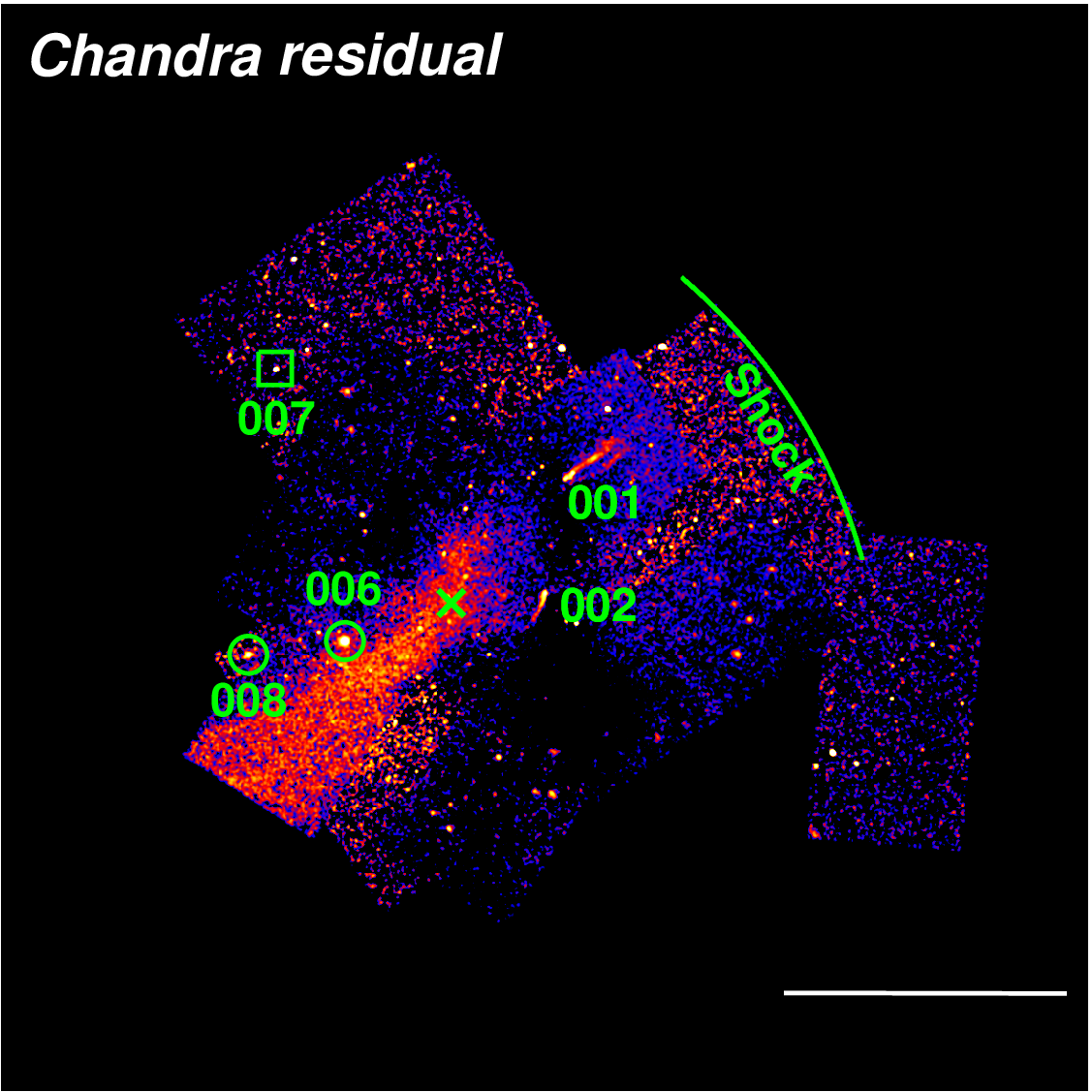}
\vspace{-0.1cm}
\caption{
{\it Top left}: \XMM\ 0.5-2 keV mosaic of A3627. The cyan contours are from ASKAP 944 MHz \citep{Koribalski2024} and show the WAT radio galaxy ESO~137-006 and the head-tail radio galaxy ESO~137-007 marked by a green box; dashed cyan contours depict the arc-shaped radio filaments resembling a `smoke ring' beyond the radio tail of ESO~137-007. Green circles mark two BCGs: ESO~137-006 and ESO~137-008. The green `X' labels the X-ray peak. Green arrows illustrate the direction of motion of two RPS galaxies with X-ray tails: ESO~137-001 and ESO~137-002. We find a merger shock front (marked by the green arc) indicated by a surface brightness edge on the NW side of the cluster. 
{\it Top right}: \XMM\ residual image with the best-fit ICM model subtracted. Prominent residual features include the CC remnant and the X-ray tails of ESO~137-001 or ESO~137-002. The peak of the CC remnant is marked by a black `+'. The radio lobes of ESO~137-006 likely excavate two X-ray cavities indicated by the X-ray decrements. 
{\it Bottom left}: \cha\ 0.5-2 keV mosaic marked with bright X-ray sources and tails.
{\it Bottom right}: \cha\ residual image with the same best-fit ICM model from the \XMM\ image subtracted. The CC remnant with X-ray cavities and the X-ray tails are also remarkable. The same bar at the bottom right of all panels shows 15$^{\prime}$/300 kpc. 
}
\label{fig:img}
\end{center}
\end{figure*}

We process the \cha\ and \XMM\ data listed in Table~\ref{t:obs}, following the procedures of \cite{Ge2018}.
We reduce the \cha\ ACIS observations with the \cha\ Interactive Analysis of Observation (CIAO; v4.17) and calibration database (CALDB; v4.12.2), and the \XMM\ EPIC data with the Extended Source Analysis Software (ESAS), as integrated into the \XMM\ Science Analysis System (SAS; v17.0.0). 
The reduced background-subtracted, exposure-corrected, and smoothed mosaics of A3627 are shown in Fig.~\ref{fig:img}.
The shock detection is mainly attributed to our new deep \XMM\ pointing (obsid = 0920830101).

The spectral analysis is also exclusively performed on the \XMM\ data.
We utilize the double-background subtraction method to subtract the instrumental non-X-ray background and X-ray backgrounds separately (\citealt{Ge2021b}).
Because the X-ray emission in mosaics (Fig.~\ref{fig:img}) is dominated by A3627, we select an off-cluster region to model the contribution of the X-ray backgrounds.
The off-cluster region (obsid = 0601741001 in Table~\ref{t:obs}, which targets an AGN IGR J16024-6107 with RA = 16:01:47.8, DEC = -61:08:55.2) lies 92 arcmin (1.8 Mpc) away from the A3627 center and is free from diffuse contamination.
We use the solar abundance table of \cite{Asplund09} in the spectral fits.
We adopt an absorption column density of $N_{\rm H} = (2.04\pm0.25) \times 10^{21}\ {\rm cm}^{-2}$ based on the spectral fitting of the X-ray peak (RA = 16:14:16.9,
DEC = -60:52:25.9) of A3627. There is a very bright X-ray point source $\sim 23^{\prime}$ to the northwest of the cluster peak (Fig.~\ref{fig:img}), it is a Seyfert 1 galaxy (WKK 6092 with RA = 16:11:51.8, DEC = -60:37:54.9) within the cluster (e.g. \citealt{Boehringer1996}). We also constrain the absorption column density based on the spectral fitting from this Seyfert galaxy of $N_{\rm H} = (1.74\pm0.17) \times 10^{21}\ {\rm cm}^{-2}$, which is consistent with one from the cluster X-ray peak accounting for the errors.

\begin{table}
\centering
\setlength{\tabcolsep}{4pt}
\caption{\chandra\ and \XMM\ Observations}
\begin{tabular}{lccc}
\hline
ObsID & PI & Date Observed & Exposure (ks)$^a$ \\
\hline
\cha\ \\
4956  & Jones & 2004-06-14 & 14.5/14.5 \\
4957  & Jones & 2004-06-14  & 14.1/14.1 \\
4958  & Jones & 2004-06-15 & 14.1/14.1 \\
8178  & Sun & 2007-07-08 & 57.4/57.2 \\
9518  & Sun & 2008-06-13 & 140.0/139.8 \\
12950  & Sun & 2011-01-10 & 89.9/89.5 \\
26789  & Nicholl & 2023-08-07 & 50.9/50.7 \\
\XMM\ \\
0208010101 & Jones & 2004-08-11 & 16.7(15.0)/12.1(7.8)\\
0208010201 & Jones & 2004-08-12 & 18.5(16.8)/13.1(9.2)\\
0204250101 & Sakelliou & 2004-09-19 & 20.4(16.9)/5.9(2.2)\\
0550451101 & Bassani & 2009-02-18 & 17.6(16.0)/11.8(5.7)\\
0601741001$^b$ & Bassani & 2009-09-09 & 23.6(22.0)/22.6(18.3)\\
0920830101 & Sun/Ge & 2024-02-24 & 89.6(93.6)/84.6(47.9)  \\
\hline
\end{tabular}
\begin{tablenotes}
\item
$^a$: Total/clean exposure time, for \XMM, the exposure times are different from MOS and pn (in brackets).
$^b$: This is an off-cluster observation to model the X-ray backgrounds.
\end{tablenotes}
\label{t:obs}
\end{table}

\section{results}
\label{sec:result}

\subsection{ICM properties}
\label{sec:icm}
We first model the global ICM distribution with a $\beta$-model \citep{1976A&A....49..137C}.
To achieve this, we extract surface brightness profiles (SBPs) centered on the X-ray peak from the \XMM\ image (Fig.~\ref{fig:img}), excluding regions of point sources and diffuse substructures. Three SBPs are extracted: an azimuthally averaged profile, as well as profiles along the major and minor axes of A3627. For the major axis profile, we adopt a position angle (PA) of $126^\circ$ from the north to the east, with opening angles of $90^\circ$ (the SBP extracted within $\pm 45^\circ$ of the major axis). The major axis region is excluded for the SBP along the minor axis. 
We then apply the $\beta$-model to fit the SBPs within 5$^{\prime}$-30$^{\prime}$ around the X-ray peak, as there is a central excess of X-ray emission associated with a substructure \citep{Boehringer1996}. 
The $\beta$-model gas distribution is given by
\begin{equation}
n_{\rm ICM}(r) =n_{\rm e0}\left(1+\frac{r^2}{r_c^2}\right)^{-\frac{3}{2}\beta},
\end{equation} 
with the derived X-ray SBP also following a $\beta$-model in the form of 
\begin{equation}   
I_{\rm X}(r) =I_0\left(1+\frac{r^2}{r_c^2}\right)^{\frac{1}{2}-3\beta}.
\end{equation}  
We apply the analytical formula given by Eq.~(3) of \cite{Ge2018} to convert the central surface brightness $I_0$ (from the $\beta$-model fitting of the SBP) to the central electron density $n_{\rm e0}$.
The resulting best-fit parameters are listed below.
\begin{equation*}  
\begin{aligned}  
& {\rm Azimuthally\ averaged:}\\
& n_{\rm e0}=(2.23\pm0.05)\times 10^{-3}{\rm\ cm^{-3}}, \\
& I_0=(6.28\pm0.09)\times 10^{-6}{\rm\ counts\ s^{-1}\ arcsec^{-2}}, \\ 
& r_c=629.7\pm22.6 {\rm\ arcsec},\ \beta=0.50\pm0.01.
\end{aligned}  
\end{equation*} 

\begin{equation*}  
\begin{aligned}  
& {\rm Along\ the\ major\ axis:}\\
& n_{\rm e0}=(2.16\pm0.05)\times 10^{-3}{\rm\ cm^{-3}}, \\
& I_0=(6.59\pm0.07)\times 10^{-6}{\rm\ counts\ s^{-1}\ arcsec^{-2}}, \\
& r_c=813.1\pm29.0 {\rm\ arcsec},\ \beta=0.59\pm0.02.
\end{aligned}  
\end{equation*}  

\begin{equation*}  
\begin{aligned}  
& {\rm Along\ the\ minor\ axis:}\\
& n_{\rm e0}=(2.25\pm0.08)\times 10^{-3}{\rm\ cm^{-3}}, \\
& I_0=(5.85\pm0.14)\times 10^{-6}{\rm\ counts\ s^{-1}\ arcsec^{-2}}, \\
& r_c=584.5\pm33.2 {\rm\ arcsec},\ \beta=0.51\pm0.02.
\end{aligned}  
\end{equation*} 

The density difference at the boundary of major and minor axes is $\sim 20\%$, which can be regarded as a systematic error due to the geometry.

The aforementioned $\beta$-model fits along the major and minor axes suggest an elliptical surface brightness distribution.
To reveal the substructures, we subtract a 2D $\beta$-model ({\tt beta2d} in {\tt SHERPA})  from both the \XMM\ and \cha\ images to produce the residual images in Fig.~\ref{fig:img}, following procedures given by \cite{Ge2020}. 
Notable residual substructures include the bright X-ray tails behind the RPS galaxies ESO~137-001 and ESO~137-002, the SE excess probably from a cool core (CC) remnant, and the NW enhancement from the shock presented in the Sec.~\ref{sec:shock}.
The SE excess and NW enhancement are also evident in the $\beta$-model subtracted residual {\em ROSAT} image \citep{Boehringer1996}.

\subsection{Merger shock}
\label{sec:shock}
\begin{figure}
\begin{center}
\centering
\includegraphics[angle=0,width=0.49\textwidth]{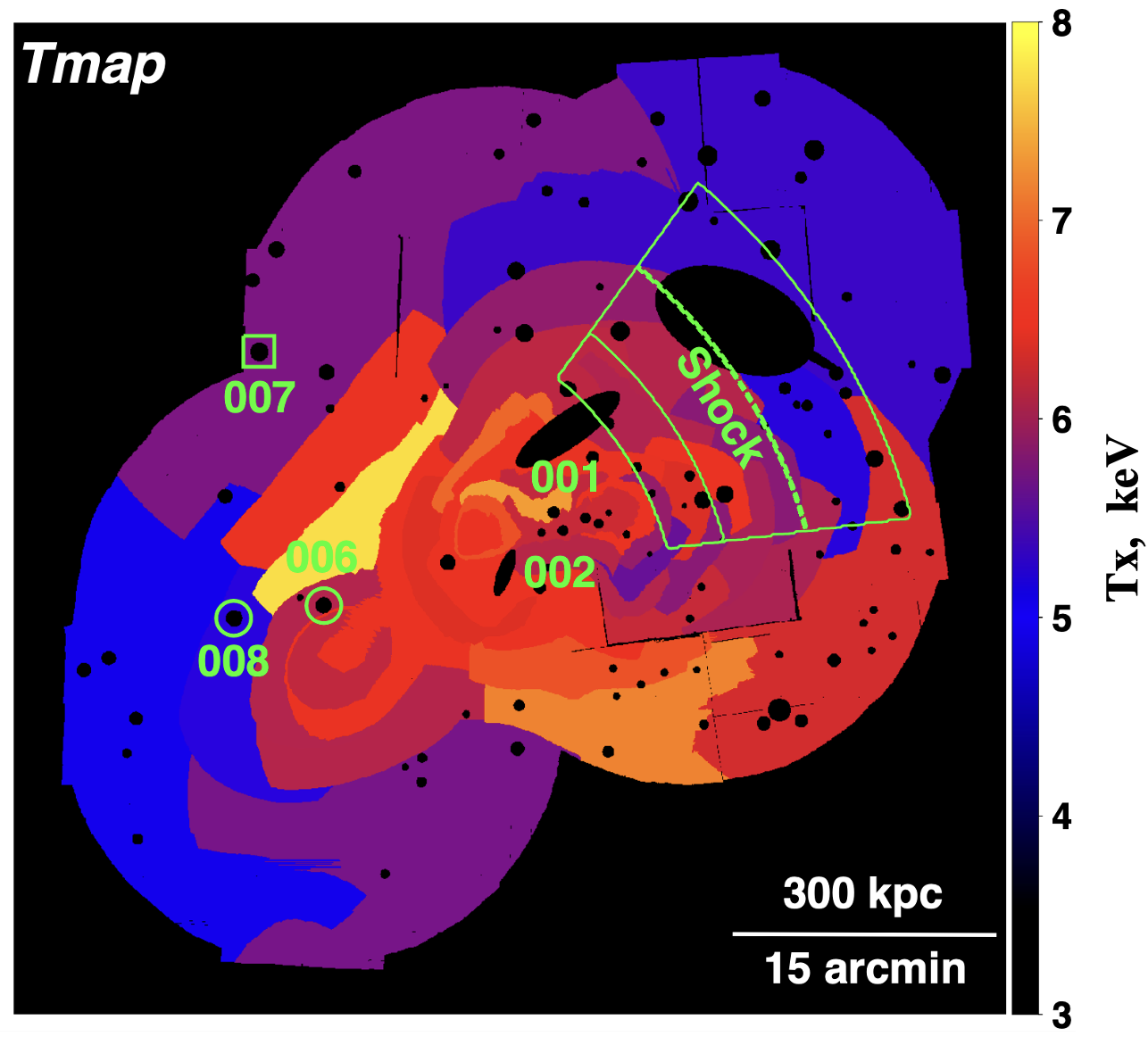}
\includegraphics[angle=0,width=0.49\textwidth]{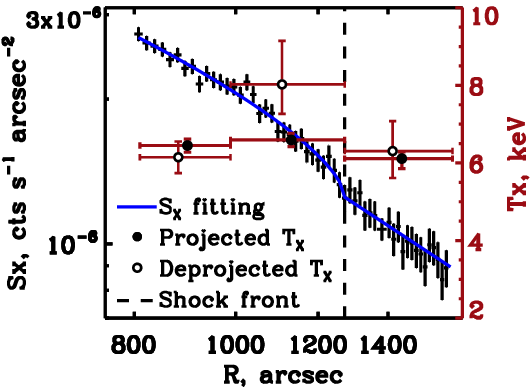}
\caption{
{\it Top}: \XMM\ temperature map of A3627 with a color bar in units of keV. Regions containing point sources and diffuse substructures are masked as black circles and ellipses. The green sector marks the extraction region for the SBP and the temperature profile shown in the bottom panel. The green dashed line marks the location of the SF. Notable galaxies are marked as in Fig.~\ref{fig:img}.
{\it Bottom}: SBP and temperature profiles near the shock. The blue solid line shows the best-fit broken power-law model. Black dots are the projected temperature, while open circles are the deprojected temperature. The vertical error bars show $1\sigma$ uncertainties. The dashed line denotes the location of the SF, where a discontinuity in the density and temperature of the ICM is observed.
}
\label{fig:rp}
\end{center}
\end{figure}

Early X-ray observations suggest that A3627 is undergoing a major merger event in the SE-NW direction as indicated by the elongated morphology and temperature gradient (\citealt{Boehringer1996,Tamura1998,Nishino2012}).
We also produce a temperature map with the Contour Binning algorithm \citep{Sanders2006} following \cite{Ge2019b} in Fig.~\ref{fig:rp}.
The \XMM\ and \cha\ observations confirm the consistent SE-NW morphological elongation and temperature gradient observed in earlier studies.
However, they have higher sensitivity and resolution to reveal more details. On the NW side of A3627, we notice a surface brightness edge in Fig.~\ref{fig:img}, and a temperature drop in the temperature map in Fig.~\ref{fig:rp}. These features are typically produced by a cluster merger shock front (SF). 
To confirm the presence of this SF and investigate its properties, we extract an SBP and a temperature profile.

The SBP is extracted from the NW of A3627 in self-similar elliptical pie annuli enclosed within the green sector shown in Fig.~\ref{fig:rp}. The shape of the ellipses matches approximately the sharp outer surface brightness edge. Fig.~\ref{fig:rp} shows the SBP; the dashed line marks the location of the surface brightness discontinuity with the changes in slope.
We fit different power-law functions beside this discontinuity (\citealt{Sarazin2016}).
The best-fit jump in gas density across the edge is $\rho_2/\rho_1=1.2\pm0.1$, where subscripts 2 and 1 denote the post-shock and pre-shock regions, respectively. We then apply the Rankine-Hugoniot jump conditions (e.g. \citealt{Nolting2019}) to derive the shock Mach number of $M_{\rho}=1.2\pm0.1$.

Additionally, we measure the temperature jump between the post-shock and pre-shock regions as shown in Fig.~\ref{fig:rp}.  The post-shock $T_X$ is $6.6\pm0.2$ keV, while the pre-shock $T_X$ is $6.1\pm0.2$ keV.
Then we adopt the ‘onion-peeling’ technique (e.g. \citealt{Ge2019b}) in elliptical shells to deproject the temperature.
The deprojected post-shock $T_X$ is $8.2\pm0.9$ keV, while the deprojected pre-shock $T_X$ is $6.3\pm0.7$ keV.
The resultant $T_X$ jump is $T_2/T_1=1.3\pm0.2$. Applying the Rankine-Hugoniot jump condition, we calculate the corresponding Mach number to be $M_T=1.3\pm0.2$.

The shock Mach numbers derived from the density and temperature jumps are consistent with each other within the errors. The small discrepancy between the two Mach numbers may be due to the projection effects (e.g. \citealt{Ge2019b}). We adopt $M_i=1.3$ as the shock Mach number of the ICM in the following discussions.
 
\section{Discussion}
\label{sec:discussion}
\subsection{Shock-bubble interactions}\label{sec:shockbubble}

The radio contours in Fig.~\ref{fig:img} show two spectacular radio galaxies. One of them is the head-tail radio galaxy ESO~137-007, which has a tail of over 500 kpc with some interesting substructures, including vortex-like features near the position of the SF.
Recent 3D MHD simulations suggest that the upstream jet of a radio galaxy may be reversed by the ICM shock, leading to essentially one-sided ``head-tail" morphologies of radio galaxies (e.g. \citealt{Jones2017}). 
We note that the green arc in Fig.~\ref{fig:img} only denotes part of the large-scale shock surface with a relatively regular shape; the actual large-scale shock front can be more extended and complex as indicated by cluster simulations, and part of the actual shock front may have overrun the radio tails. Although the shock extends beyond the green arc, it does not necessarily have the same curvature as the green arc. An oblique shock wind can also deflect and bend the jet as suggested by simulations, but the two jets may not align with each other (e.g. \citealt{Nolting2019}). The head-tail radio morphology suggests that the shock wind is nearly aligned with the jet.

The dashed cyan contours in Fig.~\ref{fig:img} show the arc-shaped radio filaments extending beyond the radio tail of ESO~137-007.
These filaments are likely a partial vortex ring structure generated by the shock.
As the shock penetrates the AGN bubble/cocoon/cavity inflated by the AGN jet, the cocoon is crushed relatively quickly, and the perimeters of the cocoon roll into a vortex ring structure resembling a `smoke ring' (e.g. \citealt{Enblin2002,Pfrommer2011,Jones2017}).
Given that the density inside the cocoon is much lower than that in the ICM, the shock velocity is greater inside the cocoon than in the surrounding ICM.
Consequently, the internal shock passes through the cocoon much faster than the external shock propagates around it.
As the shock passes through the low density cocoon, it indeed moves substantially faster through that region. This significantly distorts the shape of the shock locally, causing it to bulge outward where it passes through the cocoon material. However, the location where the shock is detected will be determined by the large-scale shock surface, which lags behind the region that encountered the cocoon. Therefore, the vortex ring may appear to be on the pre-shock side, despite having passed through the shock.
In some circumstances, the cocoon may be crushed and stripped away from the propagating jet before the shock in the external medium propagates around it (e.g. \citealt{Jones2017,Nolting2019}). This likely clarifies why the partial vortex ring structure associated with the jet of ESO~137-007 is located behind the external ICM shock in A3627. 

Based on the positions of the ICM shock and the radio ring, we can estimate the properties of the cocoon or cavity produced by the AGN jet.
From Fig.~\ref{fig:img}, the external shock is located 450 kpc away from the AGN of ESO~137-007. The internal shock has crushed the cocoon and produced the vortex ring, which is located approximately 760 kpc away from the AGN.
The rate of cocoon or cavity collapse is determined by the velocity of the original ICM-cavity contact discontinuity (CD), $v_{CD}$.
For simplicity and an order-of-magnitude estimation, we can assume that the proper motion of ESO~137-007 is relatively small compared with the shock velocity, and the length of the original jet or cocoon is relatively short compared with that of the current stretched one.
Then the velocity ratio between $v_{CD}$ and the external shock velocity in ICM $v_{si}$ is $v_{CD}/v_{si}=760/450=1.7$,  
The $v_{CD}$ obtained in the Riemann solution (e.g. \citealt{Jones2017}) is
\begin{equation}
v_{CD}=\frac{2}{\gamma_c+1}\frac{M_c^2-1}{M_c^2}v_{sc},
\label{e:vcd}
\end{equation}
where $v_{sc}=M_ca_c$ is the internal shock velocity related to the Mach number $M_c$ and sound speed $a_c$ in the cavity.
The sound speeds in the ICM and cavity are $a_i=\sqrt{\gamma_iP_i/\rho_i}$ and  $a_c=\sqrt{\gamma_cP_c/\rho_c}$, respectively.
Assuming the cavity is filled with ultra-hot thermal plasma ($\gamma_c=\gamma_i=5/3$) in pressure equilibrium with the ICM ($P_c=P_i$), we have $a_c/a_i=\sqrt{\rho_i/\rho_c}$ and then $v_{sc}/v_{si}=(M_c/M_i)\sqrt{\rho_i/\rho_c}$.
Substituting $v_{sc}$ with $v_{si}$ in Eq.~(\ref{e:vcd})
we get
\begin{equation}
\frac{v_{CD}}{v_{si}}=\frac{2}{\gamma_c+1}\frac{M_c^2-1}{M_cM_i}\sqrt{\frac{\rho_i}{\rho_c}},  
\end{equation}
then we adopt $v_{CD}/v_{si}=1.7$ and $M_i=1.3$ to have
\begin{equation}
M_c-\frac{1}{M_c}=2.9\sqrt{\delta},
\end{equation}
where $\delta=\rho_c/\rho_i$ is the density contrast between the cavity and ICM.
For a typical density contrast of $\delta=0.001-0.01$ (e.g. \citealt{Pfrommer2011,Nolting2019}), the Mach number in the cavity is $M_c=1.05-1.16$, which is also a weak shock.
Moreover, the resultant $M_c$ is insensitive to the velocity ratio $v_{CD}/v_{si}$. Therefore, even if the proper motion of ESO~137-007 and the length of the original jet cannot be ignored, although the velocity ratio $v_{CD}/v_{si}$ changes, the results remain largely unchanged.   
However, in the weak shock limit, $v_{sc}/v_{si} \sim 1/\sqrt{\delta}$ (e.g. \citealt{Jones2017}), the internal shock velocity in the cavity is much larger than the external shock velocity in the ICM due to the extreme density contrast. 
This substantial velocity difference represents a large velocity shear at the boundary between the cavity and the external medium, and this shear leads to the generation of vorticity and the vortex ring.
In the case of ESO~137-007 in A3627, the pre-shock average temperature is $T_X \sim 6$ keV, the corresponding ICM sound speed is $a_i=\sqrt{\frac{\gamma kT}{\mu m_p}}=1264{\rm\ km\ s^{-1}}$. Then the related external shock speed is $v_{si}=1643{\rm\ km\ s^{-1}}$, the cavity collapse velocity is $v_{CD}=2793{\rm\ km\ s^{-1}}$, and the internal shock velocity is $v_{sc} \gtrsim 16430 {\rm\ km\ s^{-1}}$.
The cavity collapse velocity $v_{CD}$ is between the external shock velocity $v_{si}$ and the internal shock velocity $v_{sc}$. Therefore, the cavity is crushed before the ICM shock propagates around it, as observed.

Later, after the vortex ring has been advected away, a naked jet without a cocoon remains, like the ``source C" in A2256 (e.g. \citealt{Jones2017,Nolting2019}). We also detect a merger shock near the end of the jet of ``source C" in A2256 \citep{Ge2020}, which indicates that the jet plasma cocoon has been stripped away by the shock.

The shock-bubble interactions may also explain the origin of the odd radio circles (ORCs), which are circles of radio continuum emission with low surface brightness (e.g. \citealt{Norris2021,Koribalski2021}).
The shock impact can turn the jet cocoons or lobes into vortex ring structures (e.g. \citealt{Nolting2019}). If our line of sight is near the jet or lobe axis, the vortex ring structures may resemble the ORC. 
Simulations suggest that shock reacceleration of remnant radio lobes can lead to ORC-like radio morphologies (e.g. \citealt{Shabala2024}).

\subsection{Shocks enhance RPS}\label{sec:rps}
Similar to the way an explosive shock wave strips leaves from trees, the cluster merger shock can remove gas from member galaxies.
Simulations suggest that large-scale strong shocks can significantly enhance RPS. These shocks can accelerate the ICM gas and increase its density, thereby boosting ram pressure. In some extreme cases, the ram pressure can surge by 1–3 orders of magnitude (e.g. \citealt{Li2023}). 
The signatures of RPS are notably prevalent in merging clusters that host potential merger shocks (e.g. \citealt{stroe2015,ruggiero2019,ebeling2019}).
In particular, RPS galaxies tend to be found near cluster merger shocks (e.g. \citealt{Owers2012,Ge2019b}).

The RPS galaxy ESO~137-001 is potentially in the post-shock region as indicated by the higher temperature in the temperature map (Fig.~\ref{fig:rp}), while the RPS galaxy ESO~137-002 is at a greater projected distance from the SF.
The shock may have overrun ESO~137-001, thereby enhancing the ambient ICM pressure and thus promoting the tail formation and star forming in tails (e.g. \citealt{Roediger2014}). 
Although the star formation rates ($\sim 1\ {\rm M}_\odot\ {\rm yr}^{-1}$) in ESO~137-001 and ESO~137-002 are comparable, star formation is notably more active in the tails of ESO~137-001 than in those of ESO~137-002 (e.g. \citealt{Laudari2022,Waldron2023}).
As shown in Fig.~\ref{fig:rgb}, the X-ray tails ($>80$ kpc) of ESO~137-001 are also longer and brighter than those  ($\sim 40$ kpc) of ESO~137-002 (e.g. \citealt{Sun2010,Zhang2013}).
Additionally, the geometry of mergers involving shocks and bright RPS galaxies is similar between A3627 (e.g. RPS galaxy ESO~137-001; \citealt{Sun2006,Sun2010}) and A1367 (e.g. RPS galaxy UGC~6697; \citealt{Sun2005,Ge2019b}).

If we assume that ESO~137-001 is falling into A3627 in the plane of the sky with a velocity of $\sim 1000{\rm \ km\ s}^{-1}$, and the shock of A3627 has swept through it.
We can then estimate the enhancement of ram pressure due to the shock with the standard Rankine–Hugoniot relations (e.g. \citealt{Nolting2019}).
We also adopt an ICM shock Mach number $M_i=1.3$, then the density ratio is $\rho_w/\rho_i = 1.4$, where $\rho_w$ is the density in the post-shock wind and $\rho_i$ is the density in the pre-shock ICM.
The post-shock wind velocity (e.g. \citealt{Jones2017}) is 
\begin{equation}
v_w=\frac{3}{4}\frac{M_i^2-1}{M_i}a_i,  
\end{equation}
and $a_i=1264 {\rm\ km\ s}^{-1}$ is the sound speed of the ICM in A3627. 
Then the wind velocity induced by the shock is $v_w=503{\rm\ km\ s}^{-1}$.
Because the wind velocity is in the opposite direction to the galaxy velocity $v_g$, the galaxy velocity relative to the ICM in the post-shock wind is $v_{gw}=1503 {\rm\ km\ s}^{-1}$.
The ratio of ram pressure after and before the shock is $\frac{\rho_w}{\rho_i}(\frac{v_{gw}}{v_g})^2=1.4 \times 1.5^2=3.2$. Even a weak shock can enhance the ram pressure significantly.
 Furthermore, the shock Mach numbers derived from the X-ray observations are typically underestimated due to, e.g., the projection effects (e.g. \citealt{Akamatsu2017,Zhang2019}).
Simulations of cluster formation show that shocks are actually rather complex and inhomogeneous, and that the shock strengths vary with location. 
The galaxy may have encountered a shock with a larger Mach number.
If the actual shock near ESO~137-001 has a moderate Mach number of $M_i \sim 2$, the ram pressure can be boosted by a factor of 13.
Therefore, the formation of the brightest known X-ray tail behind the cluster late-type galaxy ESO~137-001 may be attributed to the shock.

Fig.~\ref{fig:rgb} also illustrates radio tails from the synchrotron radiation produced by relativistic electrons moving within magnetic fields behind these RPS galaxies (e.g. \citealt{Koribalski2024}).
The radio tail of ESO~137-001 is spatially correlated with its X-ray tail, suggesting that most relativistic electrons originate from the host galaxy plasma stripped away by the ram pressure.
The radio tail is shorter than the X-ray tail, because its length is limited by the radiative synchrotron and inverse-Compton cooling (e.g. \citealt{Chen2020}).
The radio tail of ESO~137-002 is offset $\sim 30^\circ$ from its X-ray tail.
Unlike ESO~137-001, which is a star-forming galaxy (e.g. \citealt{Fossati2016}), ESO~137-002 hosts a Seyfert2-like AGN (e.g. \citealt{Sun2010}).
The radio tail of ESO~137-002 likely represents a nuclear jet that has been deflected or bent by the ram pressure of the wind (e.g. \citealt{Nolting2019}).

\subsection{Cause of the cool core remnant}
\label{ccr}
The global morphology of A3627 is significantly disturbed by the cluster merger, leading to an extension and distortion of its central core region.
We estimate the cooling time of the central region using an electron density $n_{\rm e0}=2.23\times 10^{-3}{\rm\ cm^{-3}}$ derived from the $\beta$-model fitting, along with a temperature $T_X=6.4$ keV and metallicity $Z=0.55\;Z_{\odot}$ obtained from the spectral fitting of a circular region (radius $\sim 3^{\prime}$) around the X-ray peak with the {\sc apec} model.
The calculated central cooling time is approximately  21.5 Gyr, determined by dividing the total thermal energy by the X-ray emissivity, expressed as $(3/2) n_{\rm tot} kT / (n_{\rm e} n_{\rm H} \Lambda$)
where the total particle density $n_{\rm tot}=1.93 n_e$, and $\Lambda$ is the cooling rate calculated from the {\sc apec} model in {\sc xspec}   (\citealt{Arnaud1996}), in the 0.01 - 100 keV band.
Meanwhile, the computed central entropy $kT/n_e^{2/3}$ is 375 keV cm$^2$.
Thus, A3627 is a typical non-CC cluster according to the criteria of, e.g. \cite{Hudson2010}. 
In fact, it is the second brightest non-CC cluster following the Coma cluster \citep{Boehringer1996}.
Residual \XMM\ and \cha\ images in Fig.~\ref{fig:img} show an SE excess, likely associated with a CC remnant as the remains of a CC after a heating event (e.g. \citealt{Rossetti2010}).
To search for the metal abundance excess that may be linked to this CC remnant, 
we extract spectra and estimate the metallicities from circular regions (radius $\sim 3^{\prime}$) centered on the X-ray peak (the `X' region in Fig.~\ref{fig:img}, yielding $0.55\pm0.12\;Z_{\odot}$) and on the remnant peak (the `+' region in Fig.~\ref{fig:img}, yielding $0.53\pm0.10\;Z_{\odot}$).
These values are then compared with the metallicity derived from a surrounding annular region (radii $\sim 8^{\prime}$–$11^{\prime}$ around both the X‑ray peak and the remnant peak), which gives $0.34\pm0.05\;Z_{\odot}$. The significant metallicity excess supports the notion that the SE excess feature is a CC remnant (e.g. \citealt{Rossetti2010}).

Observations find that cluster core properties correlate with the dynamical state or ICM morphology: CC clusters tend to be dynamically relaxed with regular morphology, while non-CC clusters are often dynamically young with disturbed morphology (e.g. \citealt{Chen2007,Pratt2010,Yuan2022}).
Simulations indicate that CCs can be disrupted by cluster mergers and heated through shock-heating and mixing (e.g. \citealt{ZuHone2011,Valdarnini2021}). 
Thus, the significant merger event in A3627 may have disrupted and heated its original CC to a CC remnant.

However, cluster mergers are not the sole cause of CC remnants, as indicated by simulations (e.g. \citealt{Rasia2015,Barnes2018}). AGNs can also heat and perturb CCs (e.g. \citealt{Ehlert2011,Liu2024}). 
Occasionally, large CC clusters experience strong AGN outbursts. During such events, powerful radio jets penetrate the inner regions of CCs, depositing substantial energy into the outer regions. The CC can be disrupted and turned into a CC remnant or corona. If the CC is completely destroyed, a non-CC may form (e.g. \citealt{Sun2009b,Liu2024}). In particular, in A3627, its BCG ESO~137-006 embeds one of the brightest coronae in clusters \citep{Sun2009b}. ESO~137-006 also hosts the WAT source PKS~1610-60, as one of the brightest radio sources \citep{Jones1996}. The radio lobes of PKS~1610-60 excavate the X-ray cavities in Fig.~\ref{fig:img}, providing compelling evidence of AGN feedback and CC disruption. 

The total energy required to excavate a cavity is equal to its enthalpy, given by
\begin{equation}
    E_{cav}=\frac{\gamma}{\gamma-1}pV,
\end{equation}
where $p$ is the pressure of the gas surrounding the cavity, $V$ is the cavity’s volume, and $\gamma$ is the ratio of specific heats of the cavity plasma; $\gamma=4/3$ for relativistic plasma and $\gamma=5/3$ for nonrelativistic plasma (e.g. \citealt{McNamara2007}). For the cavities of PKS~1610-60 filled with radio lobes, we adopt $\gamma=4/3$, and the enthalpy is $4pV$. 
For an order-of-magnitude estimation, we assume the cavity is spherical and find its radius to be $\sim 45$ kpc. The thermal pressure of the surrounding ICM, $p=n_{\rm tot}kT$, is derived with $n_{\rm tot}$ from the best-fit $\beta$-model and $T$ from the temperature map. Then the AGN feedback energy traced by the cavities is $E_{cav}=2.6\times 10^{60}$ erg.
The age of the cavity is calculated as the time required for the cavity to rise the projected distance ($R\sim 127$ kpc) from the radio core to its present location at the speed of sound, $t_{cav}=R/a_i=9.8\times 10^7$ yr (e.g. \citealt{Birza2004}). The cavity power is $P_{cav}=E_{cav}/t_{cav}=8.3\times10^{44}{\rm\ ergs\ s}^{-1}$, which represents a lower limit to the total power of the AGN, because the remaining roughly half of the AGN feedback energy is deposited into the ICM through the shock (e.g. \citealt{McNamara2007}). 
This amount of feedback energy deposited into the ICM can heat the central gas within $\sim 100$ kpc (a typical cooling radius of clusters; e.g. \citealt{Birza2004,Rafferty2006}). In the central region, the cooling through X-ray emission is $L_X=\int n_{\rm e} n_{\rm H} \Lambda dV=9.2\times10^{42}{\rm\ ergs\ s}^{-1}$, where the densities are from the best-fit $\beta$-model and the $\Lambda$ is the cooling rate calculated from the {\sc apec} model in {\sc xspec} in the 0.01 - 100 keV band. Compared with the cavity power, the cooling can be ignored. The total number of particles in the central region is $N= \int n_{\rm tot} dV=4.8\times 10^{68}$, and they can be heated with $E_{cav}\sim (3/2) N k\Delta T$. Thus, the increase in central temperature is $k\Delta T \sim 2.3$ keV per particle, which can significantly heat the central region and potentially smear out the temperature difference between the CC and the surrounding gas. Furthermore, hydrodynamical simulations show that AGN feedback and cluster mergers work together to increase the central entropy and destroy the CCs (e.g. \citealt{Chen2024,Lehle2025}).

\section{Conclusions}
\label{sec:conclusion}
We revisit the closest rich cluster, A3627, with \XMM\ and \cha\ mosaic. After subtracting the global 2D $\beta$-model of the ICM, two prominent substructures are identified: an enhancement in the NW and an excess in the SE.

The NW enhancement is related to a merger shock identified from the jumps in ICM density and temperature. 
The merger shock substantially elevates the ICM pressure and induces shocking features in the cluster galaxies.
Two extreme cases are the RPS galaxy ESO~137-001 and the head-tail radio galaxy ESO~137-007.
The shock is likely to have overrun ESO~137-001, promoting the tail formation and star forming in its tail. 
For ESO~137-007, the shock may reverse its upstream jet to a one-sided radio head-tail morphology, strip and roll the jet cocoon into a vortex ring behind the jet.

The SE excess is a CC remnant. The original CC has probably been destroyed by the violent cluster merger, and/or by the AGN feedback from ESO~137-006, which is one of the brightest radio sources. 

In summary, the interactions between the ICM and member galaxies significantly influence the evolution of both.
A3627 provides an ideal laboratory for investigating these interactions.

\begin{acknowledgments}
We thank the anonymous referee for the helpful suggestions. 
This research has made use of data and/or software provided by the High Energy Astrophysics Science Archive Research Center (HEASARC).
C.G. acknowledges support from the National Natural Science Foundation of China (Nos. 12373007, 12422302).

This research employs a list of Chandra datasets, obtained by the Chandra X-ray Observatory, contained in~\dataset[DOI: 10.25574/cdc.526]{https://doi.org/10.25574/cdc.526}.
\end{acknowledgments}

\bibliography{master}{}

@PREAMBLE{"\newcommand{\noopsort}[1]{}"}

@ARTICLE{Birza2004,
       author = {{B{\^\i}rzan}, L. and {Rafferty}, D.~A. and {McNamara}, B.~R. and {Wise}, M.~W. and {Nulsen}, P.~E.~J.},
        title = "{A Systematic Study of Radio-induced X-Ray Cavities in Clusters, Groups, and Galaxies}",
      journal = {\apj},
         year = 2004,
        month = jun,
       volume = {607},
       number = {2},
        pages = {800-809},
          doi = {10.1086/383519},
archivePrefix = {arXiv},
       eprint = {astro-ph/0402348},
 primaryClass = {astro-ph},
       adsurl = {https://ui.adsabs.harvard.edu/abs/2004ApJ...607..800B}
}

@INPROCEEDINGS{Arnaud1996,
       author = {{Arnaud}, K.~A.},
        title = "{XSPEC: The First Ten Years}",
    booktitle = {Astronomical Data Analysis Software and Systems V},
         year = 1996,
       editor = {{Jacoby}, George H. and {Barnes}, Jeannette},
       series = {Astronomical Society of the Pacific Conference Series},
       volume = {101},
        month = jan,
        pages = {17},
       adsurl = {https://ui.adsabs.harvard.edu/abs/1996ASPC..101...17A}
}

@ARTICLE{Lehle2025,
       author = {{Lehle}, Katrin and {Nelson}, Dylan and {Pillepich}, Annalisa},
        title = "{What drives cluster cool-core transformations? A population-level analysis using TNG-Cluster}",
      journal = {\aap},
         year = 2025,
        month = oct,
       volume = {702},
          eid = {A25},
        pages = {A25},
          doi = {10.1051/0004-6361/202554374},
archivePrefix = {arXiv},
       eprint = {2503.01969},
 primaryClass = {astro-ph.GA},
       adsurl = {https://ui.adsabs.harvard.edu/abs/2025A&A...702A..25L}
}

@ARTICLE{Chen2024,
       author = {{Chen}, Shuang-Shuang and {Yang}, Hsiang-Yi Karen and {Schive}, Hsi-Yu and {ZuHone}, John and {Gaspari}, Massimo},
        title = "{Cool-Core Destruction in Merging Clusters with AGN Feedback and Radiative Cooling}",
      journal = {arXiv e-prints},
         year = 2024,
        month = dec,
          eid = {arXiv:2412.13595},
        pages = {arXiv:2412.13595},
          doi = {10.48550/arXiv.2412.13595},
archivePrefix = {arXiv},
       eprint = {2412.13595},
 primaryClass = {astro-ph.CO},
       adsurl = {https://ui.adsabs.harvard.edu/abs/2024arXiv241213595C}
}

@ARTICLE{Rafferty2006,
       author = {{Rafferty}, D.~A. and {McNamara}, B.~R. and {Nulsen}, P.~E.~J. and {Wise}, M.~W.},
        title = "{The Feedback-regulated Growth of Black Holes and Bulges through Gas Accretion and Starbursts in Cluster Central Dominant Galaxies}",
      journal = {\apj},
         year = 2006,
        month = nov,
       volume = {652},
       number = {1},
        pages = {216-231},
          doi = {10.1086/507672},
archivePrefix = {arXiv},
       eprint = {astro-ph/0605323},
 primaryClass = {astro-ph},
       adsurl = {https://ui.adsabs.harvard.edu/abs/2006ApJ...652..216R}
}

@ARTICLE{McNamara2007,
       author = {{McNamara}, B.~R. and {Nulsen}, P.~E.~J.},
        title = "{Heating Hot Atmospheres with Active Galactic Nuclei}",
      journal = {\araa},
         year = 2007,
        month = sep,
       volume = {45},
       number = {1},
        pages = {117-175},
          doi = {10.1146/annurev.astro.45.051806.110625},
archivePrefix = {arXiv},
       eprint = {0709.2152},
 primaryClass = {astro-ph},
       adsurl = {https://ui.adsabs.harvard.edu/abs/2007ARA&A..45..117M}
}

@ARTICLE{Akamatsu2017,
       author = {{Akamatsu}, H. and {Mizuno}, M. and {Ota}, N. and {Zhang}, Y.-Y. and {van Weeren}, R.~J. and {Kawahara}, H. and {Fukazawa}, Y. and {Kaastra}, J.~S. and {Kawaharada}, M. and {Nakazawa}, K. and {Ohashi}, T. and {R{\"o}ttgering}, H.~J.~A. and {Takizawa}, M. and {Vink}, J. and {Zandanel}, F.},
        title = "{Suzaku observations of the merging galaxy cluster Abell 2255: The northeast radio relic}",
      journal = {\aap},
         year = 2017,
        month = apr,
       volume = {600},
          eid = {A100},
        pages = {A100},
          doi = {10.1051/0004-6361/201628400},
archivePrefix = {arXiv},
       eprint = {1612.03058},
 primaryClass = {astro-ph.HE},
       adsurl = {https://ui.adsabs.harvard.edu/abs/2017A&A...600A.100A}
}

@ARTICLE{Zhang2019,
       author = {{Zhang}, Congyao and {Churazov}, Eugene and {Forman}, William R. and {Jones}, Christine},
        title = "{Standoff distance of bow shocks in galaxy clusters as proxy for Mach number}",
      journal = {\mnras},
         year = 2019,
        month = jan,
       volume = {482},
       number = {1},
        pages = {20-29},
          doi = {10.1093/mnras/sty2501},
archivePrefix = {arXiv},
       eprint = {1808.02885},
 primaryClass = {astro-ph.GA},
       adsurl = {https://ui.adsabs.harvard.edu/abs/2019MNRAS.482...20Z}
}

@ARTICLE{Ramatsoku2019,
       author = {{Ramatsoku}, M. and {Serra}, P. and {Poggianti}, B.~M. and {Moretti}, A. and {Gullieuszik}, M. and {Bettoni}, D. and {Deb}, T. and {Fritz}, J. and {van Gorkom}, J.~H. and {Jaff{\'e}}, Y.~L. and {Tonnesen}, S. and {Verheijen}, M.~A.~W. and {Vulcani}, B. and {Hugo}, B. and {J{\'o}zsa}, G.~I.~G. and {Maccagni}, F.~M. and {Makhathini}, S. and {Ramaila}, A. and {Smirnov}, O. and {Thorat}, K.},
        title = "{GASP - XVII. H I imaging of the jellyfish galaxy JO206: gas stripping and enhanced star formation}",
      journal = {\mnras},
         year = 2019,
        month = aug,
       volume = {487},
       number = {4},
        pages = {4580-4591},
          doi = {10.1093/mnras/stz1609},
archivePrefix = {arXiv},
       eprint = {1906.03686},
 primaryClass = {astro-ph.GA},
       adsurl = {https://ui.adsabs.harvard.edu/abs/2019MNRAS.487.4580R}
}

@ARTICLE{cramer2019,
       author = {{Cramer}, W.~J. and {Kenney}, J.~D.~P. and {Sun}, M. and {Crowl}, H. and {Yagi}, M. and {J{\'a}chym}, P. and {Roediger}, E. and {Waldron}, W.},
        title = "{Spectacular Hubble Space Telescope Observations of the Coma Galaxy D100 and Star Formation in Its Ram Pressure-stripped Tail}",
      journal = {\apj},
         year = 2019,
        month = jan,
       volume = {870},
       number = {2},
          eid = {63},
        pages = {63},
          doi = {10.3847/1538-4357/aaefff},
archivePrefix = {arXiv},
       eprint = {1811.04916},
 primaryClass = {astro-ph.GA},
       adsurl = {https://ui.adsabs.harvard.edu/abs/2019ApJ...870...63C}
}

@ARTICLE{Norris2021,
       author = {{Norris}, Ray P. and {Intema}, Huib T. and {Kapi{\'n}ska}, Anna D. and {Koribalski}, B{\"a}rbel S. and {Lenc}, Emil and {Rudnick}, L. and {Alsaberi}, Rami Z.~E. and {Anderson}, Craig and {Anderson}, G.~E. and {Crawford}, E. and {Crocker}, Roland and {English}, Jayanne and {Filipovi{\'c}}, Miroslav D. and {Galvin}, Tim J. and {Hopkins}, Andrew M. and {Hurley-Walker}, Natasha and {Inoue}, Susumu and {Luken}, Kieran and {Macgregor}, Peter J. and {Manojlovi{\'c}}, Pero and {Marvil}, Josh and {O'Brien}, Andrew N. and {Park}, Laurence and {Raja}, Wasim and {Shobhana}, Devika and {Venturi}, Tiziana and {Collier}, Jordan D. and {Hale}, Catherine and {Hotan}, Aidan and {Moss}, Vanessa and {Whiting}, Matthew},
        title = "{Unexpected circular radio objects at high Galactic latitude}",
      journal = {\pasa},
         year = 2021,
        month = jan,
       volume = {38},
          eid = {e003},
        pages = {e003},
          doi = {10.1017/pasa.2020.52},
archivePrefix = {arXiv},
       eprint = {2006.14805},
 primaryClass = {astro-ph.GA},
       adsurl = {https://ui.adsabs.harvard.edu/abs/2021PASA...38....3N}
}

@ARTICLE{Koribalski2021,
       author = {{Koribalski}, B{\"a}rbel S. and {Norris}, Ray P. and {Andernach}, Heinz and {Rudnick}, Lawrence and {Shabala}, Stanislav and {Filipovi{\'c}}, Miroslav and {Lenc}, Emil},
        title = "{Discovery of a new extragalactic circular radio source with ASKAP: ORC J0102-2450}",
      journal = {\mnras},
         year = 2021,
        month = jul,
       volume = {505},
       number = {1},
        pages = {L11-L15},
          doi = {10.1093/mnrasl/slab041},
archivePrefix = {arXiv},
       eprint = {2104.13055},
 primaryClass = {astro-ph.GA},
       adsurl = {https://ui.adsabs.harvard.edu/abs/2021MNRAS.505L..11K}
}

@ARTICLE{Shabala2024,
       author = {{Shabala}, S.~S. and {Yates-Jones}, P.~M. and {Jerrim}, L.~A. and {Turner}, R.~J. and {Krause}, M.~G.~H. and {Norris}, R.~P. and {Koribalski}, B.~S. and {Filipovi{\'c}}, M. and {Rudnick}, L. and {Power}, C. and {Crocker}, R.~M.},
        title = "{Are Odd Radio Circles phoenixes of powerful radio galaxies?}",
      journal = {\pasa},
         year = 2024,
        month = jan,
       volume = {41},
          eid = {e024},
        pages = {e024},
          doi = {10.1017/pasa.2024.11},
archivePrefix = {arXiv},
       eprint = {2402.09708},
 primaryClass = {astro-ph.GA},
       adsurl = {https://ui.adsabs.harvard.edu/abs/2024PASA...41...24S}
}

@ARTICLE{Pfrommer2011,
       author = {{Pfrommer}, C. and {Jones}, T.~W.},
        title = "{Radio Galaxy NGC 1265 Unveils the Accretion Shock Onto the Perseus Galaxy Cluster}",
      journal = {\apj},
         year = 2011,
        month = mar,
       volume = {730},
       number = {1},
          eid = {22},
        pages = {22},
          doi = {10.1088/0004-637X/730/1/22},
archivePrefix = {arXiv},
       eprint = {1004.3540},
 primaryClass = {astro-ph.CO},
       adsurl = {https://ui.adsabs.harvard.edu/abs/2011ApJ...730...22P}
}

@ARTICLE{Dey2019,
       author = {{Dey}, Arjun and {Schlegel}, David J. and {Lang}, Dustin and {Blum}, Robert and {Burleigh}, Kaylan and {Fan}, Xiaohui and {Findlay}, Joseph R. and {Finkbeiner}, Doug and {Herrera}, David and {Juneau}, St{\'e}phanie and {Landriau}, Martin and {Levi}, Michael and {McGreer}, Ian and {Meisner}, Aaron and {Myers}, Adam D. and {Moustakas}, John and {Nugent}, Peter and {Patej}, Anna and {Schlafly}, Edward F. and {Walker}, Alistair R. and {Valdes}, Francisco and {Weaver}, Benjamin A. and {Y{\`e}che}, Christophe and {Zou}, Hu and {Zhou}, Xu and {Abareshi}, Behzad and {Abbott}, T.~M.~C. and {Abolfathi}, Bela and {Aguilera}, C. and {Alam}, Shadab and {Allen}, Lori and {Alvarez}, A. and {Annis}, James and {Ansarinejad}, Behzad and {Aubert}, Marie and {Beechert}, Jacqueline and {Bell}, Eric F. and {BenZvi}, Segev Y. and {Beutler}, Florian and {Bielby}, Richard M. and {Bolton}, Adam S. and {Brice{\~n}o}, C{\'e}sar and {Buckley-Geer}, Elizabeth J. and {Butler}, Karen and {Calamida}, Annalisa and {Carlberg}, Raymond G. and {Carter}, Paul and {Casas}, Ricard and {Castander}, Francisco J. and {Choi}, Yumi and {Comparat}, Johan and {Cukanovaite}, Elena and {Delubac}, Timoth{\'e}e and {DeVries}, Kaitlin and {Dey}, Sharmila and {Dhungana}, Govinda and {Dickinson}, Mark and {Ding}, Zhejie and {Donaldson}, John B. and {Duan}, Yutong and {Duckworth}, Christopher J. and {Eftekharzadeh}, Sarah and {Eisenstein}, Daniel J. and {Etourneau}, Thomas and {Fagrelius}, Parker A. and {Farihi}, Jay and {Fitzpatrick}, Mike and {Font-Ribera}, Andreu and {Fulmer}, Leah and {G{\"a}nsicke}, Boris T. and {Gaztanaga}, Enrique and {George}, Koshy and {Gerdes}, David W. and {Gontcho}, Satya Gontcho A. and {Gorgoni}, Claudio and {Green}, Gregory and {Guy}, Julien and {Harmer}, Diane and {Hernandez}, M. and {Honscheid}, Klaus and {Huang}, Lijuan Wendy and {James}, David J. and {Jannuzi}, Buell T. and {Jiang}, Linhua and {Joyce}, Richard and {Karcher}, Armin and {Karkar}, Sonia and {Kehoe}, Robert and {Kneib}, Jean-Paul and {Kueter-Young}, Andrea and {Lan}, Ting-Wen and {Lauer}, Tod R. and {Le Guillou}, Laurent and {Le Van Suu}, Auguste and {Lee}, Jae Hyeon and {Lesser}, Michael and {Perreault Levasseur}, Laurence and {Li}, Ting S. and {Mann}, Justin L. and {Marshall}, Robert and {Mart{\'\i}nez-V{\'a}zquez}, C.~E. and {Martini}, Paul and {du Mas des Bourboux}, H{\'e}lion and {McManus}, Sean and {Meier}, Tobias Gabriel and {M{\'e}nard}, Brice and {Metcalfe}, Nigel and {Mu{\~n}oz-Guti{\'e}rrez}, Andrea and {Najita}, Joan and {Napier}, Kevin and {Narayan}, Gautham and {Newman}, Jeffrey A. and {Nie}, Jundan and {Nord}, Brian and {Norman}, Dara J. and {Olsen}, Knut A.~G. and {Paat}, Anthony and {Palanque-Delabrouille}, Nathalie and {Peng}, Xiyan and {Poppett}, Claire L. and {Poremba}, Megan R. and {Prakash}, Abhishek and {Rabinowitz}, David and {Raichoor}, Anand and {Rezaie}, Mehdi and {Robertson}, A.~N. and {Roe}, Natalie A. and {Ross}, Ashley J. and {Ross}, Nicholas P. and {Rudnick}, Gregory and {Safonova}, Sasha and {Saha}, Abhijit and {S{\'a}nchez}, F. Javier and {Savary}, Elodie and {Schweiker}, Heidi and {Scott}, Adam and {Seo}, Hee-Jong and {Shan}, Huanyuan and {Silva}, David R. and {Slepian}, Zachary and {Soto}, Christian and {Sprayberry}, David and {Staten}, Ryan and {Stillman}, Coley M. and {Stupak}, Robert J. and {Summers}, David L. and {Sien Tie}, Suk and {Tirado}, H. and {Vargas-Maga{\~n}a}, Mariana and {Vivas}, A. Katherina and {Wechsler}, Risa H. and {Williams}, Doug and {Yang}, Jinyi and {Yang}, Qian and {Yapici}, Tolga and {Zaritsky}, Dennis and {Zenteno}, A. and {Zhang}, Kai and {Zhang}, Tianmeng and {Zhou}, Rongpu and {Zhou}, Zhimin},
        title = "{Overview of the DESI Legacy Imaging Surveys}",
      journal = {\aj},
         year = 2019,
        month = may,
       volume = {157},
       number = {5},
          eid = {168},
        pages = {168},
          doi = {10.3847/1538-3881/ab089d},
archivePrefix = {arXiv},
       eprint = {1804.08657},
 primaryClass = {astro-ph.IM},
       adsurl = {https://ui.adsabs.harvard.edu/abs/2019AJ....157..168D}
}

@ARTICLE{Enblin2002,
       author = {{En{\ss}lin}, T.~A. and {Br{\"u}ggen}, M.},
        title = "{On the formation of cluster radio relics}",
      journal = {\mnras},
         year = 2002,
        month = apr,
       volume = {331},
       number = {4},
        pages = {1011-1019},
          doi = {10.1046/j.1365-8711.2002.05261.x},
archivePrefix = {arXiv},
       eprint = {astro-ph/0104233},
 primaryClass = {astro-ph},
       adsurl = {https://ui.adsabs.harvard.edu/abs/2002MNRAS.331.1011E}
}

@ARTICLE{Koribalski2024,
       author = {{Koribalski}, B{\"a}rbel S. and {Duchesne}, Stefan W. and {Lenc}, Emil and {Venturi}, Tiziana and {Botteon}, Andrea and {Shabala}, Stanislav S. and {Vernstrom}, Tessa and {Carretti}, Ettore and {Norris}, Ray P. and {Anderson}, Craig and {Hopkins}, Andrew M. and {Riseley}, C.~J. and {Gupta}, Nikhel and {Velovi{\'c}}, Velibor},
        title = "{ASKAP reveals the radio tail structure of the Corkscrew Galaxy shaped by its passage through the Abell 3627 cluster}",
      journal = {\mnras},
         year = 2024,
        month = sep,
       volume = {533},
       number = {1},
        pages = {608-620},
          doi = {10.1093/mnras/stae1838},
archivePrefix = {arXiv},
       eprint = {2405.04374},
 primaryClass = {astro-ph.GA},
       adsurl = {https://ui.adsabs.harvard.edu/abs/2024MNRAS.533..608K}
}

@ARTICLE{Chen2020,
       author = {{Chen}, Hao and {Sun}, Ming and {Yagi}, Masafumi and {Bravo-Alfaro}, Hector and {Brinks}, Elias and {Kenney}, Jeffrey and {Combes}, Francoise and {Sivanandam}, Suresh and {Jachym}, Pavel and {Fossati}, Matteo and {Gavazzi}, Giuseppe and {Boselli}, Alessandro and {Nulsen}, Paul and {Sarazin}, Craig and {Ge}, Chong and {Yoshida}, Michitoshi and {Roediger}, Elke},
        title = "{The ram pressure stripped radio tails of galaxies in the Coma cluster}",
      journal = {\mnras},
         year = 2020,
        month = aug,
       volume = {496},
       number = {4},
        pages = {4654-4673},
          doi = {10.1093/mnras/staa1868},
archivePrefix = {arXiv},
       eprint = {2004.06743},
 primaryClass = {astro-ph.GA},
       adsurl = {https://ui.adsabs.harvard.edu/abs/2020MNRAS.496.4654C}
}

@ARTICLE{Laudari2022,
       author = {{Laudari}, Sunil and {J{\'a}chym}, Pavel and {Sun}, Ming and {Waldron}, Will and {Chatzikos}, Marios and {Kenney}, Jeffrey and {Luo}, Rongxin and {Nulsen}, Paul and {Sarazin}, Craig and {Combes}, Fran{\c{c}}oise and {Edge}, Tim and {Voit}, Mark and {Donahue}, Megan and {Cortese}, Luca},
        title = "{ESO 137-002: a large spiral undergoing edge-on ram-pressure stripping with little star formation in the tail}",
      journal = {\mnras},
         year = 2022,
        month = jan,
       volume = {509},
       number = {3},
        pages = {3938-3956},
          doi = {10.1093/mnras/stab3280},
archivePrefix = {arXiv},
       eprint = {2111.01821},
 primaryClass = {astro-ph.GA},
       adsurl = {https://ui.adsabs.harvard.edu/abs/2022MNRAS.509.3938L}
}

@ARTICLE{Waldron2023,
       author = {{Waldron}, William and {Sun}, Ming and {Luo}, Rongxin and {Laudari}, Sunil and {Chatzikos}, Marios and {Sivanandam}, Suresh and {Kenney}, Jeffrey D.~P. and {J{\'a}chym}, Pavel and {Voit}, G. Mark and {Donahue}, Megan and {Fossati}, Matteo},
        title = "{HST viewing of spectacular star-forming trails behind ESO 137-001}",
      journal = {\mnras},
         year = 2023,
        month = jun,
       volume = {522},
       number = {1},
        pages = {173-194},
          doi = {10.1093/mnras/stad963},
archivePrefix = {arXiv},
       eprint = {2302.07270},
 primaryClass = {astro-ph.GA},
       adsurl = {https://ui.adsabs.harvard.edu/abs/2023MNRAS.522..173W}
}

@ARTICLE{Ehlert2011,
       author = {{Ehlert}, S. and {Allen}, S.~W. and {von der Linden}, A. and {Simionescu}, A. and {Werner}, N. and {Taylor}, G.~B. and {Gentile}, G. and {Ebeling}, H. and {Allen}, M.~T. and {Applegate}, D. and {Dunn}, R.~J.~H. and {Fabian}, A.~C. and {Kelly}, P. and {Million}, E.~T. and {Morris}, R.~G. and {Sanders}, J.~S. and {Schmidt}, R.~W.},
        title = "{Extreme active galactic nucleus feedback and cool-core destruction in the X-ray luminous galaxy cluster MACS J1931.8-2634}",
      journal = {\mnras},
         year = 2011,
        month = mar,
       volume = {411},
       number = {3},
        pages = {1641-1658},
          doi = {10.1111/j.1365-2966.2010.17801.x},
archivePrefix = {arXiv},
       eprint = {1010.0253},
 primaryClass = {astro-ph.CO},
       adsurl = {https://ui.adsabs.harvard.edu/abs/2011MNRAS.411.1641E}
}

@ARTICLE{Liu2024,
       author = {{Liu}, Wenhao and {Sun}, Ming and {Voit}, G. Mark and {Lal}, Dharam Vir and {Nulsen}, Paul and {Gaspari}, Massimo and {Sarazin}, Craig and {Ehlert}, Steven and {Zheng}, Xianzhong},
        title = "{X-ray cool core remnants heated by strong radio AGN feedback}",
      journal = {\mnras},
         year = 2024,
        month = jun,
       volume = {531},
       number = {1},
        pages = {2063-2078},
          doi = {10.1093/mnras/stae1285},
archivePrefix = {arXiv},
       eprint = {2405.09738},
 primaryClass = {astro-ph.GA},
       adsurl = {https://ui.adsabs.harvard.edu/abs/2024MNRAS.531.2063L}
}

@ARTICLE{Rasia2015,
       author = {{Rasia}, E. and {Borgani}, S. and {Murante}, G. and {Planelles}, S. and {Beck}, A.~M. and {Biffi}, V. and {Ragone-Figueroa}, C. and {Granato}, G.~L. and {Steinborn}, L.~K. and {Dolag}, K.},
        title = "{Cool Core Clusters from Cosmological Simulations}",
      journal = {\apjl},
         year = 2015,
        month = nov,
       volume = {813},
       number = {1},
          eid = {L17},
        pages = {L17},
          doi = {10.1088/2041-8205/813/1/L17},
archivePrefix = {arXiv},
       eprint = {1509.04247},
 primaryClass = {astro-ph.CO},
       adsurl = {https://ui.adsabs.harvard.edu/abs/2015ApJ...813L..17R}
}

@ARTICLE{Barnes2018,
       author = {{Barnes}, David J. and {Vogelsberger}, Mark and {Kannan}, Rahul and {Marinacci}, Federico and {Weinberger}, Rainer and {Springel}, Volker and {Torrey}, Paul and {Pillepich}, Annalisa and {Nelson}, Dylan and {Pakmor}, R{\"u}diger and {Naiman}, Jill and {Hernquist}, Lars and {McDonald}, Michael},
        title = "{A census of cool-core galaxy clusters in IllustrisTNG}",
      journal = {\mnras},
         year = 2018,
        month = dec,
       volume = {481},
       number = {2},
        pages = {1809-1831},
          doi = {10.1093/mnras/sty2078},
archivePrefix = {arXiv},
       eprint = {1710.08420},
 primaryClass = {astro-ph.CO},
       adsurl = {https://ui.adsabs.harvard.edu/abs/2018MNRAS.481.1809B}
}

@ARTICLE{Valdarnini2021,
       author = {{Valdarnini}, R. and {Sarazin}, C.~L.},
        title = "{A study of cool core resiliency and entropy mixing in simulations of galaxy cluster mergers}",
      journal = {\mnras},
         year = 2021,
        month = jul,
       volume = {504},
       number = {4},
        pages = {5409-5436},
          doi = {10.1093/mnras/stab1126},
archivePrefix = {arXiv},
       eprint = {2104.08358},
 primaryClass = {astro-ph.CO},
       adsurl = {https://ui.adsabs.harvard.edu/abs/2021MNRAS.504.5409V}
}

@ARTICLE{ZuHone2011,
       author = {{ZuHone}, J.~A.},
        title = "{A Parameter Space Exploration of Galaxy Cluster Mergers. I. Gas Mixing and the Generation of Cluster Entropy}",
      journal = {\apj},
         year = 2011,
        month = feb,
       volume = {728},
       number = {1},
          eid = {54},
        pages = {54},
          doi = {10.1088/0004-637X/728/1/54},
archivePrefix = {arXiv},
       eprint = {1004.3820},
 primaryClass = {astro-ph.CO},
       adsurl = {https://ui.adsabs.harvard.edu/abs/2011ApJ...728...54Z}
}

@ARTICLE{Pratt2010,
       author = {{Pratt}, G.~W. and {Arnaud}, M. and {Piffaretti}, R. and {B{\"o}hringer}, H. and {Ponman}, T.~J. and {Croston}, J.~H. and {Voit}, G.~M. and {Borgani}, S. and {Bower}, R.~G.},
        title = "{Gas entropy in a representative sample of nearby X-ray galaxy clusters (REXCESS): relationship to gas mass fraction}",
      journal = {\aap},
         year = 2010,
        month = feb,
       volume = {511},
          eid = {A85},
        pages = {A85},
          doi = {10.1051/0004-6361/200913309},
archivePrefix = {arXiv},
       eprint = {0909.3776},
 primaryClass = {astro-ph.CO},
       adsurl = {https://ui.adsabs.harvard.edu/abs/2010A&A...511A..85P}
}

@ARTICLE{Yuan2022,
       author = {{Yuan}, Z.~S. and {Han}, J.~L. and {Wen}, Z.~L.},
        title = "{Dynamical state of galaxy clusters evaluated from X-ray images}",
      journal = {\mnras},
         year = 2022,
        month = jun,
       volume = {513},
       number = {2},
        pages = {3013-3021},
          doi = {10.1093/mnras/stac1037},
archivePrefix = {arXiv},
       eprint = {2204.02699},
 primaryClass = {astro-ph.GA},
       adsurl = {https://ui.adsabs.harvard.edu/abs/2022MNRAS.513.3013Y}
}

@ARTICLE{Chen2007,
       author = {{Chen}, Y. and {Reiprich}, T.~H. and {B{\"o}hringer}, H. and {Ikebe}, Y. and {Zhang}, Y. -Y.},
        title = "{Statistics of X-ray observables for the cooling-core and non-cooling core galaxy clusters}",
      journal = {\aap},
         year = 2007,
        month = may,
       volume = {466},
       number = {3},
        pages = {805-812},
          doi = {10.1051/0004-6361:20066471},
archivePrefix = {arXiv},
       eprint = {astro-ph/0702482},
 primaryClass = {astro-ph},
       adsurl = {https://ui.adsabs.harvard.edu/abs/2007A&A...466..805C}
}

@ARTICLE{Rossetti2010,
       author = {{Rossetti}, M. and {Molendi}, S.},
        title = "{Cool core remnants in galaxy clusters}",
      journal = {\aap},
         year = 2010,
        month = feb,
       volume = {510},
          eid = {A83},
        pages = {A83},
          doi = {10.1051/0004-6361/200913156},
archivePrefix = {arXiv},
       eprint = {0910.4900},
 primaryClass = {astro-ph.CO},
       adsurl = {https://ui.adsabs.harvard.edu/abs/2010A&A...510A..83R}
}

@ARTICLE{Hudson2010,
       author = {{Hudson}, D.~S. and {Mittal}, R. and {Reiprich}, T.~H. and {Nulsen}, P.~E.~J. and {Andernach}, H. and {Sarazin}, C.~L.},
        title = "{What is a cool-core cluster? a detailed analysis of the cores of the X-ray flux-limited HIFLUGCS cluster sample}",
      journal = {\aap},
         year = 2010,
        month = apr,
       volume = {513},
          eid = {A37},
        pages = {A37},
          doi = {10.1051/0004-6361/200912377},
archivePrefix = {arXiv},
       eprint = {0911.0409},
 primaryClass = {astro-ph.CO},
       adsurl = {https://ui.adsabs.harvard.edu/abs/2010A&A...513A..37H}
}

@ARTICLE{Nolting2019,
       author = {{Nolting}, Chris and {Jones}, T.~W. and {O'Neill}, Brian J. and {Mendygral}, P.~J.},
        title = "{Interactions between Radio Galaxies and Cluster Shocks. I. Jet Axes Aligned with Shock Normals}",
      journal = {\apj},
         year = 2019,
        month = may,
       volume = {876},
       number = {2},
          eid = {154},
        pages = {154},
          doi = {10.3847/1538-4357/ab16d6},
archivePrefix = {arXiv},
       eprint = {1904.05943},
 primaryClass = {astro-ph.HE},
       adsurl = {https://ui.adsabs.harvard.edu/abs/2019ApJ...876..154N}
}

@ARTICLE{Jones2017,
       author = {{Jones}, T.~W. and {Nolting}, Chris and {O'Neill}, B.~J. and {Mendygral}, P.~J.},
        title = "{Using collisions of AGN outflows with ICM shocks as dynamical probes}",
      journal = {Physics of Plasmas},
         year = 2017,
        month = apr,
       volume = {24},
       number = {4},
          eid = {041402},
        pages = {041402},
          doi = {10.1063/1.4978620},
archivePrefix = {arXiv},
       eprint = {1612.05700},
 primaryClass = {astro-ph.HE},
       adsurl = {https://ui.adsabs.harvard.edu/abs/2017PhPl...24d1402J}
}

@ARTICLE{Sarazin2016,
       author = {{Sarazin}, Craig L. and {Finoguenov}, Alexis and {Wik}, Daniel R. and {Clarke}, Tracy E.},
        title = "{Deep XMM-Newton Observations of the NW Radio Relic Region of Abell 3667}",
      journal = {arXiv e-prints},
         year = 2016,
        month = jun,
          eid = {arXiv:1606.07433},
        pages = {arXiv:1606.07433},
          doi = {10.48550/arXiv.1606.07433},
archivePrefix = {arXiv},
       eprint = {1606.07433},
 primaryClass = {astro-ph.HE},
       adsurl = {https://ui.adsabs.harvard.edu/abs/2016arXiv160607433S}
}

@ARTICLE{Sanders2006,
       author = {{Sanders}, J.~S.},
        title = "{Contour binning: a new technique for spatially resolved X-ray spectroscopy applied to Cassiopeia A}",
      journal = {\mnras},
         year = 2006,
        month = sep,
       volume = {371},
       number = {2},
        pages = {829-842},
          doi = {10.1111/j.1365-2966.2006.10716.x},
archivePrefix = {arXiv},
       eprint = {astro-ph/0606528},
 primaryClass = {astro-ph},
       adsurl = {https://ui.adsabs.harvard.edu/abs/2006MNRAS.371..829S}
}

@ARTICLE{Ge2021b,
       author = {{Ge}, Chong and {Sun}, Ming and {Yagi}, Masafumi and {Fossati}, Matteo and {Forman}, William and {J{\'a}chym}, Pavel and {Churazov}, Eugene and {Zhuravleva}, Irina and {Boselli}, Alessandro and {Jones}, Christine and {Ji}, Li and {Luo}, Rongxin},
        title = "{The BIG X-ray tail}",
      journal = {\mnras},
         year = 2021,
        month = nov,
       volume = {508},
       number = {1},
        pages = {L69-L73},
          doi = {10.1093/mnrasl/slab108},
archivePrefix = {arXiv},
       eprint = {2109.07964},
 primaryClass = {astro-ph.GA},
       adsurl = {https://ui.adsabs.harvard.edu/abs/2021MNRAS.508L..69G}
}

@ARTICLE{Ramatsoku2020,
       author = {{Ramatsoku}, M. and {Murgia}, M. and {Vacca}, V. and {Serra}, P. and {Makhathini}, S. and {Govoni}, F. and {Smirnov}, O. and {Andati}, L.~A.~L. and {de Blok}, E. and {J{\'o}zsa}, G.~I.~G. and {Kamphuis}, P. and {Kleiner}, D. and {Maccagni}, F.~M. and {Moln{\'a}r}, D. Cs. and {Ramaila}, A.~J.~T. and {Thorat}, K. and {White}, S.~V.},
        title = "{Collimated synchrotron threads linking the radio lobes of ESO 137-006}",
      journal = {\aap},
         year = 2020,
        month = apr,
       volume = {636},
          eid = {L1},
        pages = {L1},
          doi = {10.1051/0004-6361/202037800},
archivePrefix = {arXiv},
       eprint = {2004.03203},
 primaryClass = {astro-ph.GA},
       adsurl = {https://ui.adsabs.harvard.edu/abs/2020A&A...636L...1R}
}

@ARTICLE{Jones1996,
       author = {{Jones}, Paul A. and {McAdam}, W. Bruce},
        title = "{The head-tail and wide-angle-tail radio galaxies in cluster A3627}",
      journal = {\mnras},
         year = 1996,
        month = sep,
       volume = {282},
       number = {1},
        pages = {137-143},
          doi = {10.1093/mnras/282.1.137},
       adsurl = {https://ui.adsabs.harvard.edu/abs/1996MNRAS.282..137J}
}

@ARTICLE{Nishino2012,
       author = {{Nishino}, Sho and {Fukazawa}, Yasushi and {Hayashi}, Katsuhiro},
        title = "{Suzaku Observation of Nearby On-Going Merger Cluster Abell 3627}",
      journal = {\pasj},
         year = 2012,
        month = feb,
       volume = {64},
          eid = {16},
        pages = {16},
          doi = {10.1093/pasj/64.1.16},
       adsurl = {https://ui.adsabs.harvard.edu/abs/2012PASJ...64...16N}
}

@ARTICLE{Tamura1998,
       author = {{Tamura}, Takayuki and {Fukazawa}, Yasushi and {Kaneda}, Hidehiro and {Makishima}, Kazuo and {Tashiro}, Makoto and {Tanaka}, Yasuo and {Bohringer}, Hans},
        title = "{ASCA Observations of a Nearby and Massive Galaxy Cluster Abell 3627}",
      journal = {\pasj},
         year = 1998,
        month = apr,
       volume = {50},
        pages = {195-201},
          doi = {10.1093/pasj/50.2.195},
archivePrefix = {arXiv},
       eprint = {astro-ph/9802319},
 primaryClass = {astro-ph},
       adsurl = {https://ui.adsabs.harvard.edu/abs/1998PASJ...50..195T}
}

@ARTICLE{Boehringer1996,
       author = {{Boehringer}, H. and {Neumann}, D.~M. and {Schindler}, S. and {Kraan-Korteweg}, R.~C.},
        title = "{Abell 3627: A Nearby, X-Ray Bright, and Massive Galaxy Cluster}",
      journal = {\apj},
         year = 1996,
        month = aug,
       volume = {467},
        pages = {168},
          doi = {10.1086/177592},
archivePrefix = {arXiv},
       eprint = {astro-ph/9602140},
 primaryClass = {astro-ph},
       adsurl = {https://ui.adsabs.harvard.edu/abs/1996ApJ...467..168B}
}

@ARTICLE{Tully2014,
       author = {{Tully}, R. Brent and {Courtois}, H{\'e}l{\`e}ne and {Hoffman}, Yehuda and {Pomar{\`e}de}, Daniel},
        title = "{The Laniakea supercluster of galaxies}",
      journal = {\nat},
         year = 2014,
        month = sep,
       volume = {513},
       number = {7516},
        pages = {71-73},
          doi = {10.1038/nature13674},
archivePrefix = {arXiv},
       eprint = {1409.0880},
 primaryClass = {astro-ph.CO},
       adsurl = {https://ui.adsabs.harvard.edu/abs/2014Natur.513...71T}
}

@ARTICLE{Woudt2008,
       author = {{Woudt}, P.~A. and {Kraan-Korteweg}, R.~C. and {Lucey}, J. and {Fairall}, A.~P. and {Moore}, S.~A.~W.},
        title = "{The Norma cluster (ACO 3627) - I. A dynamical analysis of the most massive cluster in the Great Attractor}",
      journal = {\mnras},
         year = 2008,
        month = jan,
       volume = {383},
       number = {2},
        pages = {445-457},
          doi = {10.1111/j.1365-2966.2007.12571.x},
archivePrefix = {arXiv},
       eprint = {0706.2227},
 primaryClass = {astro-ph},
       adsurl = {https://ui.adsabs.harvard.edu/abs/2008MNRAS.383..445W}
}

@ARTICLE{Kraan-Korteweg1996,
       author = {{Kraan-Korteweg}, R.~C. and {Woudt}, P.~A. and {Cayatte}, V. and {Fairall}, A.~P. and {Balkowski}, C. and {Henning}, P.~A.},
        title = "{A nearby massive galaxy cluster behind the Milky Way}",
      journal = {\nat},
         year = 1996,
        month = feb,
       volume = {379},
       number = {6565},
        pages = {519-521},
          doi = {10.1038/379519a0},
       adsurl = {https://ui.adsabs.harvard.edu/abs/1996Natur.379..519K}
}

@ARTICLE{Sobral2015,
       author = {{Sobral}, David and {Stroe}, Andra and {Dawson}, William A. and {Wittman}, David and {Jee}, M. James and {R{\"o}ttgering}, Huub and {van Weeren}, Reinout J. and {Br{\"u}ggen}, Marcus},
        title = "{MC$^{2}$: boosted AGN and star formation activity in CIZA J2242.8+5301, a massive post-merger cluster at z = 0.19}",
      journal = {\mnras},
         year = 2015,
        month = jun,
       volume = {450},
       number = {1},
        pages = {630-645},
          doi = {10.1093/mnras/stv521},
archivePrefix = {arXiv},
       eprint = {1503.02076},
 primaryClass = {astro-ph.GA},
       adsurl = {https://ui.adsabs.harvard.edu/abs/2015MNRAS.450..630S}
}

@ARTICLE{Stroe2021,
       author = {{Stroe}, Andra and {Sobral}, David},
        title = "{ENISALA. II. Distinct Star Formation and Active Galactic Nucleus Activity in Merging and Relaxed Galaxy Clusters}",
      journal = {\apj},
         year = 2021,
        month = may,
       volume = {912},
       number = {1},
          eid = {55},
        pages = {55},
          doi = {10.3847/1538-4357/abe7f8},
archivePrefix = {arXiv},
       eprint = {2102.10116},
 primaryClass = {astro-ph.GA},
       adsurl = {https://ui.adsabs.harvard.edu/abs/2021ApJ...912...55S}
}

@ARTICLE{Umeda2004,
       author = {{Umeda}, Kazuyoshi and {Yagi}, Masafumi and {Yamada}, Sanae F. and {Taniguchi}, Yoshiaki and {Shioya}, Yasuhiro and {Murayama}, Takashi and {Nagao}, Tohru and {Ajiki}, Masaru and {Fujita}, Shinobu S. and {Komiyama}, Yutaka and {Okamura}, Sadanori and {Shimasaku}, Kazuhiro},
        title = "{The H{\ensuremath{\alpha}} Luminosity Function of the Galaxy Cluster A521 at z = 0.25}",
      journal = {\apj},
         year = 2004,
        month = feb,
       volume = {601},
       number = {2},
        pages = {805-812},
          doi = {10.1086/380555},
archivePrefix = {arXiv},
       eprint = {astro-ph/0310212},
 primaryClass = {astro-ph},
       adsurl = {https://ui.adsabs.harvard.edu/abs/2004ApJ...601..805U}
}

@ARTICLE{Bekki2010,
       author = {{Bekki}, Kenji and {Owers}, Matt S. and {Couch}, Warrick J.},
        title = "{Synchronized Formation of Starburst and Post-starburst Galaxies in Merging Clusters of Galaxies}",
      journal = {\apjl},
         year = 2010,
        month = jul,
       volume = {718},
       number = {1},
        pages = {L27-L31},
          doi = {10.1088/2041-8205/718/1/L27},
archivePrefix = {arXiv},
       eprint = {1006.3399},
 primaryClass = {astro-ph.CO},
       adsurl = {https://ui.adsabs.harvard.edu/abs/2010ApJ...718L..27B}
}

@ARTICLE{Owers2012,
       author = {{Owers}, Matt S. and {Couch}, Warrick J. and {Nulsen}, Paul E.~J. and {Randall}, Scott W.},
        title = "{Shocking Tails in the Major Merger Abell 2744}",
      journal = {\apjl},
         year = 2012,
        month = may,
       volume = {750},
       number = {1},
          eid = {L23},
        pages = {L23},
          doi = {10.1088/2041-8205/750/1/L23},
archivePrefix = {arXiv},
       eprint = {1204.1052},
 primaryClass = {astro-ph.CO},
       adsurl = {https://ui.adsabs.harvard.edu/abs/2012ApJ...750L..23O}
}

@ARTICLE{Ge2019b,
       author = {{Ge}, Chong and {Sun}, Ming and {Liu}, Ruo-Yu and {Rudnick}, Lawrence and {Sarazin}, Craig and {Forman}, William and {Jones}, Christine and {Chen}, Hao and {Liu}, Wenhao and {Yagi}, Masafumi and {Boselli}, Alessandro and {Fossati}, Matteo and {Gavazzi}, Giuseppe},
        title = "{A merger shock in Abell 1367}",
      journal = {\mnras},
         year = 2019,
        month = jun,
       volume = {486},
       number = {1},
        pages = {L36-L40},
          doi = {10.1093/mnrasl/slz049},
archivePrefix = {arXiv},
       eprint = {1904.08803},
 primaryClass = {astro-ph.GA},
       adsurl = {https://ui.adsabs.harvard.edu/abs/2019MNRAS.486L..36G}
}

@ARTICLE{ebeling2019,
       author = {{Ebeling}, Harald and {Kalita}, Boris S.},
        title = "{Jellyfish: Ram Pressure Stripping As a Diagnostic Tool in Studies of Cluster Collisions}",
      journal = {\apj},
         year = 2019,
        month = sep,
       volume = {882},
       number = {2},
          eid = {127},
        pages = {127},
          doi = {10.3847/1538-4357/ab35d6},
archivePrefix = {arXiv},
       eprint = {1907.12781},
 primaryClass = {astro-ph.GA},
       adsurl = {https://ui.adsabs.harvard.edu/abs/2019ApJ...882..127E}
}

@ARTICLE{ruggiero2019,
       author = {{Ruggiero}, Rafael and {Machado}, Rubens E.~G. and {Roman-Oliveira}, Fernanda V. and {Chies-Santos}, Ana L. and {Lima Neto}, Gast{\~a}o B. and {Doubrawa}, Lia and {Rodr{\'\i}guez del Pino}, Bruno},
        title = "{Galaxy cluster mergers as triggers for the formation of jellyfish galaxies: case study of the A901/2 system}",
      journal = {\mnras},
         year = 2019,
        month = mar,
       volume = {484},
       number = {1},
        pages = {906-914},
          doi = {10.1093/mnras/sty3422},
archivePrefix = {arXiv},
       eprint = {1812.05611},
 primaryClass = {astro-ph.GA},
       adsurl = {https://ui.adsabs.harvard.edu/abs/2019MNRAS.484..906R}
}

@ARTICLE{stroe2015,
       author = {{Stroe}, Andra and {Sobral}, David and {Dawson}, William and {Jee}, M. James and {Hoekstra}, Henk and {Wittman}, David and {van Weeren}, Reinout J. and {Br{\"u}ggen}, Marcus and {R{\"o}ttgering}, Huub J.~A.},
        title = "{The rise and fall of star formation in z {\ensuremath{\sim}} 0.2 merging galaxy clusters}",
      journal = {\mnras},
         year = 2015,
        month = jun,
       volume = {450},
       number = {1},
        pages = {646-665},
          doi = {10.1093/mnras/stu2519},
archivePrefix = {arXiv},
       eprint = {1410.2891},
 primaryClass = {astro-ph.GA},
       adsurl = {https://ui.adsabs.harvard.edu/abs/2015MNRAS.450..646S}
}

@ARTICLE{Li2023,
       author = {{Li}, Hao and {Wang}, Huiyuan and {Mo}, H.~J. and {Wang}, Yuan and {Luo}, Xiong and {Li}, Renjie},
        title = "{Shock-induced Stripping of the Satellite Interstellar and Circumgalactic Medium in IllustrisTNG Clusters at Z   0}",
      journal = {\apj},
         year = 2023,
        month = jan,
       volume = {942},
       number = {1},
          eid = {44},
        pages = {44},
          doi = {10.3847/1538-4357/aca7bd},
       adsurl = {https://ui.adsabs.harvard.edu/abs/2023ApJ...942...44L}
}

@ARTICLE{Sun2010,
   author = {{Sun}, M. and {Donahue}, M. and {Roediger}, E. and {Nulsen}, P.~E.~J. and 
	{Voit}, G.~M. and {Sarazin}, C. and {Forman}, W. and {Jones}, C.
	},
    title = "{Spectacular X-ray Tails, Intracluster Star Formation, and ULXs in A3627}",
  journal = {\apj},
archivePrefix = "arXiv",
   eprint = {0910.0853},
     year = 2010,
    month = jan,
   volume = 708,
    pages = {946-964},
      doi = {10.1088/0004-637X/708/2/946},
   adsurl = {http://adsabs.harvard.edu/abs/2010ApJ...708..946S}
}

@ARTICLE{Sun2009b,
   author = {{Sun}, M.},
    title = "{Every BCG with a Strong Radio Agn has an X-Ray Cool Core: Is the Cool Core-Noncool Core Dichotomy Too Simple?}",
  journal = {\apj},
archivePrefix = "arXiv",
   eprint = {0904.2006},
 primaryClass = "astro-ph.CO",
     year = 2009,
    month = oct,
   volume = 704,
    pages = {1586-1604},
      doi = {10.1088/0004-637X/704/2/1586},
   adsurl = {http://adsabs.harvard.edu/abs/2009ApJ...704.1586S}
}

@ARTICLE{Sun2006,
   author = {{Sun}, M. and {Jones}, C. and {Forman}, W. and {Nulsen}, P.~E.~J. and 
	{Donahue}, M. and {Voit}, G.~M.},
    title = "{A 70 Kiloparsec X-Ray Tail in the Cluster A3627}",
  journal = {\apjl},
   eprint = {astro-ph/0511516},
     year = 2006,
    month = feb,
   volume = 637,
    pages = {L81-L84},
      doi = {10.1086/500590},
   adsurl = {http://adsabs.harvard.edu/abs/2006ApJ...637L..81S}
}

@ARTICLE{Gunn1972,
   author = {{Gunn}, J.~E. and {Gott}, III, J.~R.},
    title = "{On the Infall of Matter Into Clusters of Galaxies and Some Effects on Their Evolution}",
  journal = {\apj},
     year = 1972,
    month = aug,
   volume = 176,
    pages = {1},
      doi = {10.1086/151605},
   adsurl = {http://adsabs.harvard.edu/abs/1972ApJ...176....1G}
}
\bibliographystyle{aasjournalv7}

\end{document}